\documentclass[12pt]{article}
\usepackage[utf8]{inputenc}
\usepackage{amsmath,amsfonts,amssymb}
\usepackage{graphicx}
\usepackage{cancel,slashed}
\usepackage{xcolor}
\usepackage[numbers, elide, compress]{natbib}
\usepackage[colorlinks=true, citecolor=blue, urlcolor=blue, linkcolor=blue, breaklinks=true,pdfpagelabels=false ]{hyperref}
\usepackage[symbol]{footmisc}
\usepackage{mathtools}
\usepackage{hyperref}
\newcommand\undermat[2]{
	\makebox[0.5pt][l]{$\smash{\underbrace{\phantom{%
					\begin{matrix}#2\end{matrix}}}_{ \let\scriptstyle\textstyle\text{\large $#1$}}}$}#2}
\newcommand\overmat[2]{
	\makebox[-1pt][l]{$\smash{\overbrace{\phantom{%
					\begin{matrix}#2\end{matrix}}}^{ \let\scriptstyle\textstyle\text{\large $#1$}}}$}#2}    

\newcommand{\Dfbd}{\mathord{\buildrel{\lower3pt\hbox{$\scriptscriptstyle\leftrightarrow$}}\over {D}_{\mu}}}

\usepackage[vcentermath]{youngtab}

\newcommand{\beq}{\begin {equation}}  
\newcommand{\eeq}{\end   {equation}} 
\newcommand{\bea}{\begin {eqnarray}} 
\newcommand{\eea}{\end   {eqnarray}}  
\newcommand{\baa}{\begin {array}   } 
\newcommand{\eaa}{\end   {array}   } 
\newcommand{\bit}{\begin {itemize} }
\newcommand{\eit}{\end   {itemize} }
\newcommand{\mc}[1]{\mathcal{#1}}

\newcommand\snowmass{\begin{center}\rule[-0.2in]{\textwidth}{0.01in}\\\rule{\textwidth}{0.01in}\\
\vskip 0.1in Submitted to the  Proceedings of the US Community Study\\ 
on the Future of Particle Physics (Snowmass 2021)\\ 
\rule{\textwidth}{0.01in}\\\rule[+0.2in]{\textwidth}{0.01in} \end{center}}

\def\equationautorefname~#1\null{Eq.\,(#1)\null}
\def\sectionautorefname~#1\null{Sec.\,#1\null}
\def\subsectionautorefname~#1\null{Sec.\,#1\null}
\def\figureautorefname~#1\null{Fig.\,#1\null}
\def\appendixautorefname~#1\null{App.\,#1\null}

\makeatletter
\def\section{\@startsection {section}{1}{\z@}{+3.0ex plus +1ex minus
  +.2ex}{2.3ex plus .2ex}{\large\bf\boldmath}}
\def\subsection{\@startsection{subsection}{2}{\z@}{+2.5ex plus +1ex
minus +.2ex}{1.5ex plus .2ex}{\normalsize\bf\boldmath}}
\def\subsubsection{\@startsection{subsubsection}{3}{\z@}{+3.25ex plus
 +1ex minus +.2ex}{1.5ex plus .2ex}{\normalsize\it}}
\makeatother

\usepackage[textwidth=16.5cm, textheight=22cm]{geometry}

\graphicspath{{.}{plots/}}

\begin{document}
\thispagestyle{empty}

\snowmass 



\vspace{1cm}

\begin{center}

{\Large\bf
Theoretical developments in the SMEFT at dimension-8 and beyond
\par}

\vspace*{3.5em}

\renewcommand{\thefootnote}{\fnsymbol{footnote}}
{\large
Simone~Alioli$^{1}$, Radja~Boughezal$^{2}$, Weiguang Cao$^{3}$, Mikael Chala$^{4}$, \'Alvaro D\'iaz-Carmona$^{4}$, Supratim Das Bakshi$^{4}$, Gauthier~Durieux$^{5}$, Luk\'{a}\v{s}~Gr\'{a}f$^{6}$, Guilherme Guedes$^{7}$, Brian Quinn~Henning$^{8}$, Teppei~Kitahara$^{9}$, Hao-Lin~Li$^{10}$, Xiaochuan~Lu$^{11}$, Camila S.~Machado$^{12}$, Adam~Martin$^{13}$, Tom~Melia$^{14}$\footnote{Corresponding author: \href{mailto:tom.melia@ipmu.jp}{tom.melia@ipmu.jp}}, Emanuele~Mereghetti$^{15}$, Hitoshi~Murayama$^{16}$, Christopher W.~Murphy, Jasper Roosmale Nepveu$^{17}$, Sridip Pal$^{18}$, Frank~Petriello$^{19}$\footnote{Corresponding author: \href{mailto:f-petriello@northwestern.edu}{f-petriello@northwestern.edu}}, Yael Shadmi$^{20}$, Jing Shu$^{21}$, Yaniv Weiss$^{22}$, Ming-Lei Xiao$^{19}$, Jiang-Hao~Yu$^{21}$
\par}

\vspace*{1cm}

\def\Northwestern{Department of Physics \& Astronomy, Northwestern University, Evanston, IL 60208, USA}
\def\Argonne{High Energy Physics Division, Argonne National Laboratory, Argonne, IL 60439, USA; }
\def\Bicocca{Dipartimento di Fisica, Universita' degli Studi di Milano-Bicocca, Milan, Italy}
\def\INFNmib{Istituto Nazionale di Fisica Nucleare, Sezione di Milano-Bicocca, Milan, Italy}
\def\ND{Department of Physics, University of Notre Dame, South Bend, IN 46556 USA}
\newcommand{\technion}{Physics Department, Technion--Israel Institute of Technology, Technion city, Haifa 3200003, Israel}
\def\Geneva{D\'{e}partment de Physique Th\'{e}orique, Universit\'{e} de Gen\`{e}ve, 24 quai Ernest-Ansermet, 1211 Gen\`{e}ve 4, Switzerland}

{\sl
$^1$ \Bicocca;\INFNmib\\[1ex]
$^2$ \Argonne\\[1ex]
$^{3}$ Kavli Institute for the Physics and Mathematics of the
  Universe (WPI), University of Tokyo Institutes for Advanced Study, University of Tokyo,
  Kashiwa 277-8583, Japan;
  Department of Physics, Graduate School of Science, The University of Tokyo, Tokyo 113-0033, Japan\\[1ex]
$^{4}$ Departamento de F\'isica Te\'orica y del Cosmos, Universidad de Granada, E-18071 Granada, Spain \\[1ex]
$^5$ CERN, Theoretical Physics Department, Geneva 23 CH-1211, Switzerland\\[1ex]
$^6$ Department of Physics, University of California, Berkeley, CA 94720, USA;
Department of Physics, University of California, San Diego, CA 92093, USA \\[1ex]
$^7$ Departamento de F\'isica Te\'orica y del Cosmos, Universidad de Granada, E-18071 Granada, Spain;  Laborat\'orio de Instrumenta\c cao e F\'isica Experimental de Part\'iculas, Departamento de
F\'isica da Universidade do Minho, Campus de Gualtar, 4710-057 Braga, Portugal\\[1ex] 
$^8$ Theoretical Particle Physics Laboratory (LPTP), Institute of Physics, EPFL, Lausanne, Switzerland\\[1ex]
$^9$ Institute for Advanced Research  \& Kobayashi-Maskawa Institute for the Origin of Particles and the Universe, 
Nagoya University,  Nagoya 464-8602, Japan\\[1ex]
$^{10}$ Centre for Cosmology, Particle Physics and Phenomenology (CP3), Universite Catholique de Louvain,\\
Chem. du Cyclotron 2, 1348, Louvain-la-neuve, Belgium\\[1ex]
$^{11}$ Institute for Fundamental Science, University of Oregon, Eugene, OR 97403, USA \\[1ex]
$^{12}$ {Deutsches Elektronen-Synchrotron DESY, Notkestr. 85, 22607 Hamburg, Germany}\\[1ex]
$^{13}$ \ND\\[1ex]
$^{14}$ Kavli Institute for the Physics and Mathematics of the
  Universe (WPI), University of Tokyo Institutes for Advanced Study, University of Tokyo, Kashiwa 277-8583, Japan\\[1ex]
$^{15}$ Theoretical Division, Los Alamos National Laboratory, Los Alamos, NM 87545, USA\\[1ex]
$^{16}$ Kavli Institute for the Physics and Mathematics of the
  Universe (WPI), University of Tokyo Institutes for Advanced Study, University of Tokyo,
  Kashiwa 277-8583, Japan;
Department of Physics, University of California, Berkeley, CA 94720, USA;
Ernest Orlando Lawrence Berkeley National Laboratory, University of California, Berkeley, CA 94720, USA\\[1ex]
$^{17}$ Humboldt-Universit\"at zu Berlin,\\ Institut f\"ur Physik, D-12489 Berlin,
Germany\\[1ex]
$^{18}$ School of Natural Sciences, Institute for Advanced Study, One Einstein Drive, Princeton, NJ 08540, U.S.A.\\[1ex]
$^{19}$ \Argonne\Northwestern{} \\[1ex]
$^{20}$ \technion\\[1ex]
$^{21}$ CAS Key Laboratory of Theoretical Physics, Insitute of Theoretical Physics,
Chinese Academy of Sciences, Beijing 100190, China;
School of Physical Sciences, University of Chinese Academy of Sciences, Beijing 100049, China;
Center for High Energy Physics, Peking University, Beijing 100871, China;
School of Fundamental Physics and Mathematical Sciences, Hangzhou Institute for Advanced Study, University of Chinese Academy of Sciences, Hangzhou 310024, China;
International Center for Theoretical Physics Asia-Pacific, Beijing/Hanzhou, China
 \\[1ex]
$^{22}$\technion \\[1ex]
}
\thispagestyle{empty}
\end{center}

\renewcommand{\thefootnote}{\arabic{footnote}}
\vspace*{1.5cm}
\setcounter{page}{0}
\setcounter{footnote}{0}

\begin{abstract}
In this contribution to the Snowmass 2021 process we review theoretical developments in the Standard Model Effective Field Theory (SMEFT) with a focus on effects at the dimension-8 level and beyond. We review the theoretical advances that led to the complete construction of the operator bases for the dimension-8 and dimension-9 SMEFT Lagrangians. We discuss the possibility of obtaining all-orders results in the $1/\Lambda$ expansion for certain SMEFT observables as well as the current status of renormalization group running and implications for positivity, and briefly present the on-shell approach to constructing SMEFT amplitudes. Finally we present several new phenomenological effects that first arise at dimension-8 and discuss the impact of these terms on experimental analyses.
\end{abstract}


\section{Introduction} \label{sec:intro}

The Standard Model (SM) has so far been remarkably successful in describing all data coming from both low-energy experiments and
high-energy colliders.  Although the search for new particles is continuing, it is becoming increasingly important to search for potentially small and subtle indirect signatures of new physics.  A convenient theoretical framework for performing such searches when only the SM particles are known is the SM effective field theory (SMEFT) which contains higher-dimensional operators formed from SM fields.   The SMEFT is an expansion in an energy scale $\Lambda$ at which the effective theory breaks down and new fields must be added to the Lagrangian.  The leading dimension-6 operators characterizing lepton-number conserving deviations from the SM have been classified for some time now~\cite{Buchmuller:1985jz, Arzt:1994gp, Grzadkowski:2010es}.

A complete list of the independent dimension-7 SMEFT operators has been worked out \cite{Lehman:2014jma}. Less is known about terms at dimension-8 and beyond.  The number of operators at each order in the expansion has been determined~\cite{Lehman:2015via, Lehman:2015coa, Henning:2015alf}, and ideas and tools have been developed towards systematically constructing these operators~\cite{Henning:2017fpj, Hays:2018zze, Henning:2019mcv, Henning:2019enq}.  Recently the complete dimension-8 SMEFT basis was constructed via brute force in~\cite{Murphy:2020rsh} and was systematically generated in~\cite{Li:2020gnx} and the complete dimension-9 SMEFT basis was constructed as well~\cite{Li:2020xlh, Liao:2020jmn}.
More recently a general procedure, implemented in a Mathematica package ABC4EFT~\cite{Li:2022tec}, has been proposed to construct the independent and complete SMEFT operator bases up to any mass dimension.

It is our goal in this contribution to review the status of the SMEFT
at dimension-8 and beyond, and motivate future studies on this topic.  We will discuss the advances that have led
to a complete construction of the operator basis at dimension-8, and
the counting of operators to all orders in $1/\Lambda$.  We will
present examples where all-orders results in the $1/\Lambda$ expansion
have been obtained.  In several cases novel phenomenological
consequences appear first at dimension-8 or beyond.  We will review such
examples in this article.

\section{Operator counting and basis structure beyond dimension-6} \label{sec:opcount}

The Hilbert series approach has been established to systematically enumerate operators in phenomenological EFTs that are subject to redundancies (symmetry group, integration by parts, equation of motion) \cite{Benvenuti:2006qr, Feng:2007ur, Gray:2008yu, Jenkins:2009dy, Hanany:2010vu, Henning:2015daa},
and was applied to the SMEFT in~\cite{Lehman:2015via, Lehman:2015coa, Henning:2015alf}.
Hilbert series are akin to partition or generating functions, containing detailed information about \emph{both} the number of operators \emph{and} the structure of the operator basis with a given field content. 
A breakdown of operators into those that are parity ($P$) even and $P$-odd can be systematically accounted for using Hilbert series methods~\cite{Henning:2017fpj}. In the white paper we will review these techniques and how to further account for charge conjugation ($C$) \cite{Graf:2020yxt}, so as e.g.\ to enable systematic identification of $CP$-odd operators at dimension-8 and above, which could have particularly striking signatures.

The concept of a Hilbert series for EFTs is very straightforward: it is \textit{defined} to be an object which counts the number of independent operators. For example, we may wish to count the number of operators of a given mass dimension, in which case the Hilbert series is
\begin{equation}
    H = \sum_{k} c_k q^k,
\end{equation}
where $c_k$ is \textit{defined} to be the number of independent operators of mass dimension $k$, while $q\in \mathbb{C}$ is just a complex number (often called a ``spurion''). For example, in the SMEFT $c_5 = 2$ (the Weinberg operator $(LH)^2$ and its Hermitian conjugate) while $c_6 = 84$ (again, Hermitian conjugate operators are counted separately). It is often very helpful to refine the definition of the Hilbert series to include detailed information about composition of operators,
\begin{equation}
    H = \sum_{r_1\dots r_n}\sum_s c_{\mathbf{r}s} \phi_1^{r_1}\phi_2^{r_2}\cdots \phi_n^{r_n} \mathcal{D}^s,
\end{equation}
where $c_{\mathbf{r}s}\equiv c_{r_1\dots r_n s}$ is defined to be the number of independent operators composed of $r_1$ fields of type $\phi_1$, $r_2$ fields of type $\phi_2$, $\dots$, and $s$ (covariant) derivatives, while $\phi_i,\mathcal{D}\in \mathbb{C}$ are spurions for the field content and derivatives. For example, in the SMEFT there are two independent operators composed of two $H$ fields, two $H^\dagger$ fields, and two derivatives (\textit{e.g.} $(\partial_{\mu} |H|^2)^2$ and $(H^{\dagger}\Dfbd H)^2$) and hence $c_{H^2H^{\dagger 2}\mathcal{D}^2} =2$.

Hilbert series techniques provide a way of computing the coefficients $c_{\mathbf{r}s}$, \textit{i.e.} of computing the number of independent operators. Here ``independent'' means that the operators give distinct physical contributions to scattering amplitudes, and hence the rules for ``independent'' derive directly from the $S$-matrix. In particular, scattering amplitudes have all particles on-shell and obey momentum conservation, as well as being Lorentz invariant and invariant under potential internal and/or gauge symmetries. Operators in a Lagrangian are in position space, $\mathcal{O}(x)$, in which case the position space avatars of ``on-shell'' and ``momentum conservation'' are respectively equations of motion and integration by parts~\cite{Henning:2015daa,Henning:2017fpj}.

\begin{figure}[htb]
\begin{centering}
\includegraphics[width=0.9\textwidth]{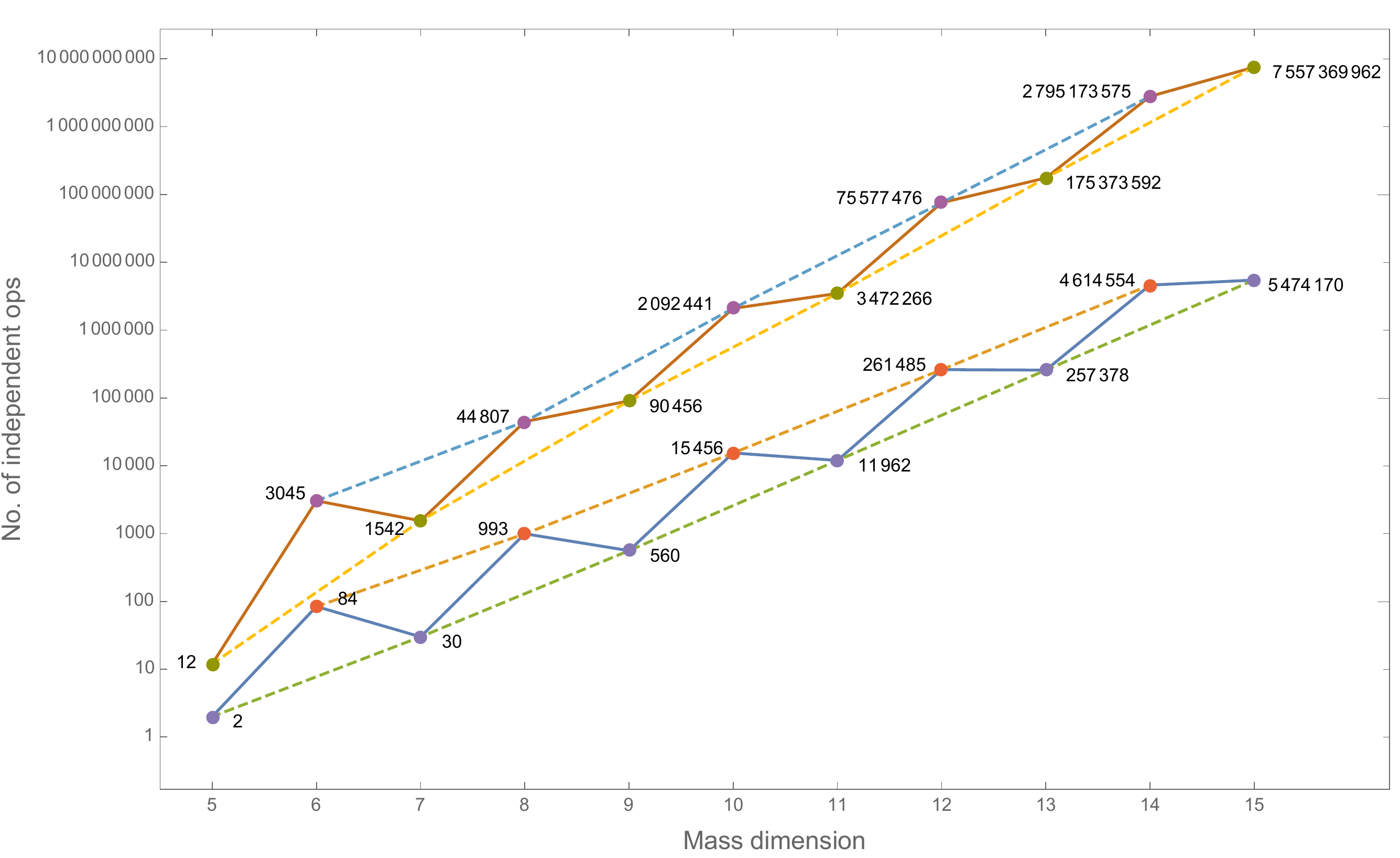}
\caption{The number of independent SMEFT operators up to dimension 15~\cite{Henning:2015alf}. Points joined by the lower (upper) solid line are for one (three) generations of fermions. Dashed lines are to guide the eye to the growth of the even and odd dimension operators.}
\label{fig:growth}
\end{centering}
\end{figure}

A series of works~\cite{Lehman:2015via,Lehman:2015coa,Henning:2015daa,Henning:2015alf,Henning:2017fpj} determined how to systematically enumerate operators (equivalently, contact terms in amplitudes~\cite{Henning:2017fpj}) accounting for redundancies associated with symmetries, equations of motion, and integration by parts. A summary of this calculation is provided at the end of this section.
Applying these tools to the SMEFT, Fig.~\ref{fig:growth} plots the number of independent operators evaluated for one and three generations of fermions up to dimension 15.\footnote{Asymptotic formulae for these curves (that agree well even at low mass dimension) have been obtained using the analytic techniques in \cite{Melia:2020pzd}. The growth of the number of operators in the SMEFT, $\rho(\Delta)$, is, for $N_g$ generations of fermions,
$$
\rho(\Delta)\underset{\Delta\to\infty}{\sim}  \mathcal{N}\exp \left(\frac{2\pi  \sqrt{2}}{3}  \sqrt[4]{7 N_g+\frac{112}{15}}\Delta ^{3/4} -\frac{\pi  \left(5 N_g+32\right)}{4 \sqrt{2} \sqrt[4]{7 N_g+\frac{112}{15}}}\Delta^{1/4}+28 \zeta '(-2)\right)\,,
$$
where $\mathcal{N}$ is given by
$$
\mathcal{N}=\frac{27783 \left(\frac{7}{5}\right)^{3/8} 3^{5/8} \pi ^{10} \left(15 N_g+16\right){}^{27/8} }{1024000 \sqrt[4]{2} \Delta ^{55/8} \left(N_g+3\right){}^4 \left(2 N_g+5\right){}^{3/2} \sqrt{10 N_g+3}}\,.
$$
}

As well as systematically enumerating EFT operators (with field content information), Hilbert series contain important information about the structure of the operator basis, and thus the structure of EFT. This is also relevant to operator construction at dimension-8 and beyond, which we review briefly.

Ref~\cite{Henning:2017fpj} established the connection between $S$-matrices (i.e. amplitudes, or more specifically contact interaction contributions to amplitudes), operator bases and partition functions, and how the Hilbert series acts to tie together these ideas. Algorithms were given to obtain basis functions (composed of kinematic scattering variables); these were used to obtain results at all-orders in the EFT expansion for four- and five-point EFT amplitudes (or, equivalently, all-order operator basis construction).\footnote{See Sec 5 of~\cite{Henning:2017fpj}; it is interesting the extent to which the all-order five-point results are  considerably more involved owing to the more non-trivial amplitude kinematics.}  The (all-order) Hilbert series functions are an important input to this; in this sense,  as laid out in~\cite{Henning:2017fpj}, Hilbert series can be seen as containing analytic information on scattering amplitudes in EFT.

Another key idea introduced in~\cite{Henning:2017fpj} was the use of conformal representation theory in organizing the operator bases for a class of EFTs  that includes the SMEFT (relativistic EFTs, with linearly realized symmetry groups). This is not surprising, as an EFT in this case is a small deformation of the free theory, which enjoys conformal symmetry. The Hilbert series, properly weighted, is simply the partition function of the free theory. Under certain assumptions (of weak coupling), this partition function can even capture asymptotic (high temperature) behavior of the interacting theory~\cite{Cao:2021euf}. Asymptotic (i.e. dimension $\to\infty$) analytic behavior of the SMEFT has been studied in~\cite{Melia:2020pzd}.

A deeper structure of the operator bases for theories such as the SMEFT can be seen in spinor-helicity variables; Refs.~\cite{Henning:2019mcv,Henning:2019enq} elucidated an underlying $SL(2,C)\times U(N)$ action---dubbed `conformal-helicity duality'\footnote{With the `conformal' referring to the $SL(2,C) \simeq SO(3,1) \subset SU(2,2)\simeq SO(4,2)$ action and a generalized `helicity' referring to the $U(N)$ action, the representation theory of which determine each other (the `duality').}---that governs EFT operator bases. This action can be leveraged to systematically construct amplitudes/operators using the representation theory (e.g. Young tableau) of the $U(N)$~\cite{Henning:2019enq}.

Another manifestation of the above mentioned  role that conformal symmetry plays in the structure of EFT operator bases can be seen through the appearance of selection rules in the anomalous dimension matrix of the theory: certain entries are zero only for operator bases consisting of conformal primary operators~\cite{Cao:2021cdt}. 

Ref~\cite{Cao:2021cdt} also exemplified the use of Hilbert series in conjunction with off-shell techniques to calculate quantum corrections in EFT.  Indeed, state-of-the-art calculations of quantum corrections (e.g. in terms of loop order) are often in practice achieved with traditional off-shell methodology. It was shown how Hilbert series can also be utilized to analyze the  basis structure of off-shell correlation functions. This is one way in which Hilbert series and operator basis structure is relevant to beyond-leading-order studies of the SMEFT.

Another way in which the above is pertinent to loop calculations is down to the fact that the Hilbert series and operator construction technologies introduced in Ref~\cite{Henning:2017fpj} work in $d$ spacetime dimensions (where a spinor-helicity approach does not in general exist), which is relevant to calculations that employ dimensional regularization. In this direction, evanescent operators (that vanish in integer dimensions, but can nevertheless leave an imprint through loop corrections) can be studied using the above techniques; systematic enumeration of a particular class of evanescent operators using Hilbert series has been presented \cite{Cao:2021cdt}. This particular class of operators appears at very high mass dimension in the SMEFT; it would be interesting to see if other classes of evanescent operators in the SMEFT, such as those that appear at mass dimension 6 and beyond, can be analyzed systematically with Hilbert series.

We succinctly summarize how to compute the Hilbert series; see \cite{Henning:2017fpj, Graf:2020yxt} for detailed elaborations. Up to corrections that only represent relevant operators (i.e. operators with dimension less or equal to four), the Hilbert series $H$ can be computed as
\begin{equation}
H (\phi, p) = \int\text{d}\mu_\text{Internal}^{}(y) \int\text{d}\mu_\text{Spacetime}^{}(x)\, \frac{1}{P(p, x)}\, Z(\phi, p, x, y) \,.
\label{eqn:H}
\end{equation}
Here the integrals $\int\text{d}\mu_\text{Internal}^{}(y)$ and $\int\text{d}\mu_\text{Spacetime}^{}(x)$ are Haar measure integrals respectively over the internal gauge symmetry groups and the Lorentz symmetry group. The factor
\begin{equation}
\frac{1}{P(p,x)}=\left(1 - p x_1\right) \left(1 - p x_1^{-1}\right) \left(1 - p x_2\right) \left(1 - p x_2^{-1}\right) \notag
\end{equation} 
accounts for the integration by parts redundancies. The integrand $Z(\phi, p, x, y)$ is a graded character of the representation generated by all fields' single particle modules, whose specific expression is
\begin{equation}
Z(\phi, p, x, y) = \text{PE} \left[ \sum_i \phi_i\, \chi_i^\text{Spacetime}(p, x)\, \chi_i^\text{Internal}(y) \right] \,.
\label{eqn:Zexpression}
\end{equation}
Here $\text{PE}$ stands for the plethystic exponential; $\chi_i^\text{Spacetime}(p, x)$ and $\chi_i^\text{Internal}(y)$ are the characters of the spacetime and internal symmetry representations of the single particle module formed by the field $\phi_i$ (and its descendants), where the equation of motion redundancies are removed.

The discrete symmetry parity (or charge conjugation) is an outer automorphism of the corresponding spacetime (or internal) symmetry group. It hence \emph{extends} the symmetry group into disconnected branches. Specifically, parity $P$ extends $SO(4)$ into $O(4) = SO(4) \rtimes \mathcal{P} = \left\{O_+(4), O_-(4)\right\}$ and charge conjugation extends $SU(N)$ into $\widetilde{SU}(N) \equiv SU(N)\rtimes\mathcal{C} = \left\{\widetilde{SU}_+(N), \widetilde{SU}_-(N)\right\}$. When these discrete symmetries are involved, one could apply Eq.~\eqref{eqn:H} to compute the Hilbert series on each of the branches
\begin{subequations}\label{eqn:HBranches}
\begin{align}
H^{C^+ P^+}(\phi, p) &\equiv \int\text{d}\mu_{\widetilde{SU}_+(N)}\left(y\right)         \int\text{d}\mu_{O_+(4)}\left(x\right)\,         \frac{1}{P_+\left(p, x\right)}\, Z^{C^+ P^+}\left(\phi, p, x, y\right) \,, \\[8pt]
H^{C^+ P^-}(\phi, p) &\equiv \int\text{d}\mu_{\widetilde{SU}_+(N)}\left(y\right)         \int\text{d}\mu_{O_-(4)}\left(\tilde{x}\right)\, \frac{1}{P_-\left(p, \tilde{x}\right)}\, Z^{C^+ P^-}\left(\phi, p, \tilde{x}, y\right) \,, \\[8pt]
H^{C^- P^+}(\phi, p) &\equiv \int\text{d}\mu_{\widetilde{SU}_-(N)}\left(\tilde{y}\right) \int\text{d}\mu_{O_+(4)}\left(x\right)\,         \frac{1}{P_+\left(p, x\right)}\, Z^{C^- P^+}\left(\phi, p, x, \tilde{y}\right) \,, \\[8pt]
H^{C^- P^-}(\phi, p) &\equiv \int\text{d}\mu_{\widetilde{SU}_-(N)}\left(\tilde{y}\right) \int\text{d}\mu_{O_-(4)}\left(\tilde{x}\right)\, \frac{1}{P_-\left(p, \tilde{x}\right)}\, Z^{C^- P^-}\left(\phi, p, \tilde{x}, \tilde{y}\right) \,,
\end{align}
\end{subequations}
and then combine them properly to obtain the following symmetric cases of the Hilbert series
\begin{subequations}\label{eqn:HcasesBasic}
\begin{alignat}{1}
H^\text{total}                     &= H^{C^+ P^+} \,, \\[8pt]
H^{C\text{-even}}                &= \frac12 \left( H^{C^+ P^+} + H^{C^- P^+} \right) \,, \\[8pt]
H^{P\text{-even}}                &= \frac12 \left( H^{C^+ P^+} + H^{C^+ P^-} \right) \,, \label{eqn:Peven} \\[8pt]
H^{C\text{-even}\,P\text{-even}} &= \frac14 \left( H^{C^+ P^+} + H^{C^+ P^-} + H^{C^- P^+} + H^{C^- P^-} \right) \,.
\end{alignat}
\end{subequations}
These Hilbert series tell us about the breakdown of operators into parity even and odd, and/or charge conjugation even and odd cases.

\section{Construction of the dimension-8 operator basis} \label{sec:dim8}

A prerequisite to studying the phenomenology of dimension-8 operators in a consistent fashion is the construction of a complete basis of dimension-8 operators.
In this section we will review the construction of the dimension-8 SMEFT basis with an emphasis on aspects of the construction that are most germane to bases at $d > 6$.
Historically, brute force was the only method to construct an operator basis.
This is the approach utilized in~\cite{Murphy:2020rsh} and~\cite{Liao:2020jmn} to generate bases of dimension-8 and -9 operators, respectively.
The brute force approach is reviewed in~\ref{sec:dim8_BF}.
A new development is the systematic approach of Refs.~\cite{Li:2020gnx, Li:2020xlh, Li:2022tec}, which can generate bases of operators to any mass dimension.
This new, systematic approach based on Young Tensors is reviewed in~\ref{sec:dim8_YT}.

Note that these methodologies can be applied to other EFTs as well.
For example, Ref.~\cite{Murphy:2020cly} used brute force to construct a basis of dimension-8 operators in the Low Energy Effective Field Theory below the Electroweak Scale (LEFT).
Ref.~\cite{Li:2020tsi} constructed LEFT operators bases through dimension-9 using the Young Tensor approach of~\cite{Li:2020gnx, Li:2020xlh, Li:2022tec}, and Ref.~\cite{Li:2021tsq} also used this systematic approach to construct bases of operators through dimension-9 in the SMEFT extended with sterile neutrinos.

\subsection{Brute Force Approach} \label{sec:dim8_BF}

Using brute force to construct an operator basis generally becomes increasingly difficult as the mass dimension, $d$, increases.
Firstly, as reviewed in Sec.~\ref{sec:opcount}, the number of operators grows exponentially with $d$.
Secondly, the most difficult cases are operators with derivatives and/or repeated fields, and these entities become more common as $d$ increases.
In this subsection we briefly sketch out how the aforementioned difficulties were overcome in Ref.~\cite{Murphy:2020rsh}.

The two complications arising in operators with derivatives are redundancies due to the equations of motion (EOM) and integration by parts (IBP).
The equations of motion can neatly be accounted for by only retaining the highest weight Lorentz representations of derivative operators~\cite{Lehman:2015via, Lehman:2015coa}.
For example, two derivatives acting on the scalar Higgs field, $H$, can be decomposed into four representations of the Lorentz group $G_L = SU(2)_l \times SU(2)_r$
\begin{equation}
\label{eq:higgs2der_reps}
    D^2H \sim (0, 0) \oplus (0, 1) \oplus (1, 0) \oplus (1, 1) .
\end{equation}
It is only the rightmost representation in~\eqref{eq:higgs2der_reps} that needs to be retained in the basis.
More generally, for $n$ derivatives acting on a scalar, $\phi$, a fermion, $\psi$, or a field strength $X$ we only need to retain  the following representations in the basis
\begin{align}
    D^n \phi &\sim (D^n H)_{(a_1 \ldots a_n), (\dot a_1 \ldots \dot a_n)} , \nonumber \\
    D^n \psi_L &\sim (D^n \psi_L)_{(a_1 \ldots a_n a_{n+1}), (\dot a_1 \ldots \dot a_n)} , \nonumber \\
    D^n X_R &\sim (D^n X_R)_{(a_1 \ldots a_n), (\dot a_1 \ldots \dot a_n \dot a_{n+1} \dot a_{n+2})} ,
\end{align}
where $a_i$ and $\dot a_i$ are fundamental indices of the left and right components of the Lorentz group, $G_L$, and round brackets, $(\ldots)$, represent symmetrization of the indices.
Redundancies due to integration by parts can be handled using the method developed in Ref.~\cite{Hays:2018zze}.
It is perhaps most easily illustrated with an example.
Consider the field content $\bar l$, $e$, $H$, $B_L$ along with two derivatives where $l$ and $e$ are left- and right-handed leptons, respectively, and $B$ is the hypercharge field strength.
There are four viable candidate dimension-8 operators with this field content after eliminating operators that can be reduced via the equation of motion as discussed above~\cite{Murphy:2020rsh}
\begin{align}
\label{eq:lqHB_cand}
x_1 &= (D \bar l)_{a, (\dot a \dot c)} e_{\dot d} (D H)_{b, \dot b} B_{(c d)} \epsilon^{a c} \epsilon^{b d} \epsilon^{\dot a \dot d} \epsilon^{\dot c \dot b}, \nonumber \\
x_2 &= \bar l_{\dot c} (D e)_{a, (\dot a \dot d)} (D H)_{b, \dot b} B_{(c d)} \epsilon^{a c} \epsilon^{b d} \epsilon^{\dot a \dot c} \epsilon^{\dot d \dot b}, \nonumber \\
x_3 &= (D \bar l)_{a, (\dot a \dot c)} (D e)_{b, (\dot b \dot d)} H B_{(c d)} \epsilon^{a c} \epsilon^{b d} \epsilon^{\dot a \dot b} \epsilon^{\dot c \dot d}, \nonumber \\
x_4 &= \bar l_{\dot c} e_{\dot d} (D^2 H)_{(a b), (\dot a \dot b)} B_{(c d)} \epsilon^{a c} \epsilon^{b d} \epsilon^{\dot a \dot c} \epsilon^{\dot b \dot d},
\end{align} 
To check for IBP redundancies we need to construct all the independent objects with the same field content but one fewer derivative that transform as $(\tfrac{1}{2}, \tfrac{1}{2})$ under the Lorentz group.
In this example, there are three such objects.
\begin{align}
\label{eq:lqHB_ibp}
y_1 &= (D \bar l)_{a, (\dot a \dot c)} e_{\dot d} H B_{(c d)} \epsilon^{a c} \epsilon^{\dot a \dot d}, \nonumber \\
y_2 &= \bar l_{\dot c} (D e)_{a, (\dot a \dot d)} H B_{(c d)} \epsilon^{a c} \epsilon^{\dot a \dot c}, \nonumber \\
y_3 &= \bar l_{\dot c} e_{\dot d} (D H)_{a, \dot a} B_{(c d)} \tfrac{1}{2} \epsilon^{a c} (\epsilon^{\dot a \dot c} + \epsilon^{\dot a \dot d}) . 
\end{align}
Candidate IBP constraints are then given by $D y_i = 0$.
However, some of these conditions may be related, which can be determined after some linear algebra.
In the example at hand, it turns out all three equations, $D y_i = 0$, are linearly independent, and we therefore have $(N_x = 4) - (N_y = 3) = 1$ operator with field content $leB_L HD^2$.
Accounting for flavor is trivial in this example as the fermion fields are not repeated.
In particular, for $n_g$ generations of fermions we simply have $n_g^2$ operators in this example.

\begin{figure}
\begin{centering}
 \includegraphics[width=0.6\textwidth]{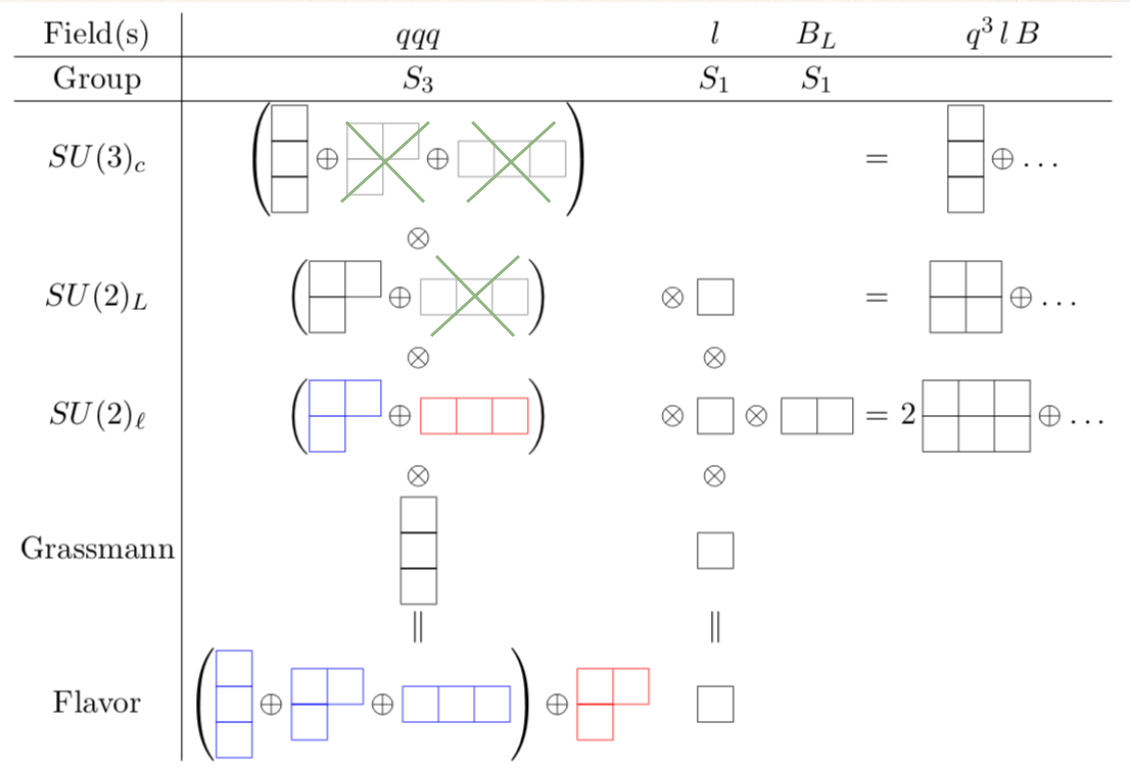}
 \caption{Computing the flavor representations for the field content $q^3 l B_L$. See the text for details.}\label{fig:qqqlB_flavorreps}
\end{centering}
\end{figure}

The process of counting operators is made non-trivial when there are repeated fields in the operator.
To handle these cases we use the method developed in Ref.~\cite{Fonseca:2019yya}.
Here, the permutation group of $n$ objects, $S_n$, plays a central role.
This method is again perhaps best highlighted with an example.
Consider dimension-8 operators with field content $q^3 l B_L$ where $q$ is the left-handed quark doublet.
Figure~\ref{fig:qqqlB_flavorreps} summarizes the procedure pictorially, and we proceed to describe it in detail in what follows.
To start, for each field in the operator, work out its representations under the gauge and Lorentz groups.
This is trivial for non-repeated fields.
Here we only need to consider non-Abelian groups as invariance under Abelian groups can be checked arithmetically.
In Fig.~\ref{fig:qqqlB_flavorreps} only the $SU(2)_l$ piece of $G_L$ is considered as all the fields in the example are left-handed.
Next multiply the rows together to determine which representations of the repeated fields lead to gauge and Lorentz invariant operators.
Representations that fail this step are crossed out with green Xs in Fig.~\ref{fig:qqqlB_flavorreps}.
It is at this point that the permutation group, $S_n$, enters.
Treat all the surviving gauge and Lorentz representations as reps.~of $S_n$.
That is, the $SU(3)_c$ singlet becomes a $\{1, 1, 1\}$ under $S_3$, the $SU(2)$ doublets becomes $\{2, 1\}$ under $S_3$, and the quartet of $SU(2)_l$ becomes a $\{3\}$ under $S_3$.
The fourth row in Fig.~\ref{fig:qqqlB_flavorreps}, labeled Grassmanian, enforces the antisymmetry of fermionic wavefunction or the symmetry of bosonic wavefunction.
Then, to get the flavor representations multiply the columns together.
Fig.~\ref{fig:qqqlB_flavorreps} is color coded such that the blue and red flavor reps.~have different Lorentz representations.
The operators themselves are
\begin{align}
    {\color{blue} Q_{lq^3B}^{(1)}} &= \epsilon_{\alpha\beta\gamma} \epsilon_{mn} \epsilon_{jk} (q_p^{m\alpha} C q_r^{j\beta}) (q_s^{k\gamma} C \sigma^{\mu\nu} l_t^n) B_{\mu\nu} , \nonumber \\
    {\color{red} Q_{lq^3B}^{(2)}} &= \epsilon_{\alpha\beta\gamma} \epsilon_{mn} \epsilon_{jk} (q_p^{m\alpha} C \sigma^{\mu\nu} q_r^{j\beta}) (q_s^{k\gamma} C l_t^n) B_{\mu\nu} ,
\end{align}
where the color coding matches that of Fig.~\ref{fig:qqqlB_flavorreps}.
The terms $Q_{lq^3B}^{(1)}$ and $Q_{lq^3B}^{(2)}$ contain $\tfrac{1}{3} n_l n_q (2 n_q^2 + 1)$ and $\tfrac{1}{3} n_l n_q (n_q^2 - 1)$ operators, respectively.
For three generations of fermions, $n_l = n_q = 3$, this yields 57 and 24 operators, respectively.

There is an ambiguity in this example: given a contraction of Lorentz indices, how should the $SU(2)_L$ gauge indices be contracted.
The origin of the ambiguity is $\{2, 1\}$ is a two-dimensional rep.~of $S_3$, and its manifestation is a redundant operator.
A priori, $Q_{lq^3B}^{(3)} = \epsilon_{\alpha\beta\gamma} \epsilon_{mj} \epsilon_{kn} (q_p^{m\alpha} C q_r^{j\beta}) (q_s^{k\gamma} C \sigma^{\mu\nu} l_t^n) B_{\mu\nu}$ is equally valid candidate to be the ``blue'' term in Fig.~\ref{fig:qqqlB_flavorreps}.
However, $Q_{lq^3B}^{(1)}$ and $Q_{lq^3B}^{(3)}$ are related~\cite{Abbott:1980zj}
\begin{equation}
    - Q_{\substack{lq^3B \\ prst}}^{(3)} = Q_{\substack{lq^3B \\ prst}}^{(1)} + Q_{\substack{lq^3B \\ rpst}}^{(1)} .
\end{equation}
The $p \leftrightarrow r$ symmetry of $Q_{lq^3B}^{(3)}$ does not allow for the antisymmetric $\{1, 1, 1\}$ rep.~of $S_3$, whereas all three of the blue flavor reps.~in Fig.~\ref{fig:qqqlB_flavorreps} are allowed by $Q_{lq^3B}^{(1)}$ indicating that it should be included in the basis.

\subsection{Young Tensor Approach} \label{sec:dim8_YT}

In Refs.~\cite{Li:2020gnx,Li:2020zfq,Li:2022tec,Li:2020xlh}, a systematic approach based on the applications of Young tableaux of symmetric groups and gauge groups is demonstrated for obtaining the explicit forms of a complete operator basis including the detailed contraction patterns for gauge and Lorentz indices, while the subtlety of the repeated fields is carefully tackled. The full results therein for dimension 8 and 9 SMEFT are generated by an automated Mathematica package ABC4EFT~\cite{Li:2022tec}, which could also generate higher dimensional SMEFT operators basis up to any mass dimension. Due to the advantage of the Young tensor approach, it is possible to systematically convert any operator into this basis via the ABC4EFT code. 

The $i$-th building block is the field with covariant derivatives $D^{w_i}\Psi_{i}$ which belongs to the reducible representations of the $SL(2,\mathbb{C})$ group and can be decomposed as a direct sum of the following irreducible representations for $\Psi_{i}$ of a irreducible representation $(j_l,j_r)$,
\begin{equation}\label{eq:LorDecom}
    D^{w_i} \Psi_i \in \left(j_{l}+\frac{w_i}{2}, j_{r}+\frac{w_i}{2}\right) \oplus \text{lower weights}.
\end{equation}
It can be shown that the ``lower weights'' in the above decomposition must contain EOM of the field or the covariant derivative commutator $[D,D]$, and by convention we can eliminate them in a \emph{type} of operators with particular constituting fields $\Psi_i$ and a fixed total number of derivatives. These terms are always understood to be converted to other types and counted therein. Since the algorithm enumerates operator bases type by type, it is sufficient to retain only the highest weights in our building blocks with the spinor indices totally symmetrized which can be expressed as:
\begin{equation}\label{eq:highestweight}
 (D^{w_i} \Psi_i)^{(\dot{\alpha}_i^1\dots \dot{\alpha}_i^{{2j_r+w_i}})}_{(\alpha_i^{1}\dots \alpha_i^{2j_l+w_i})}\sim (D^w \Psi_i)^{(\dot{\alpha}_i)^{2j_r+w_i}}_{(\alpha_i)^{2j_l+w_i}}.
\end{equation}
On the right-hand side, we abbreviate the indices $\dot{\alpha}_i$ and $\alpha_i$ to a power form without specific superscripts indicating their equivalence due to the symmetrization. 
These symmetrized spinor indices make perfect correspondence with the helicity spinor variables in the on-shell amplitudes, thus the building blocks actually make the operator basis one-to-one corresponds to an on-shell amplitude basis \cite{Ma:2019gtx,Shadmi:2018xan,Li:2020zfq}.
At the same time, the field building blocks are representations of gauge groups as well, and the gauge indices of the $i$-th field can be collectively denoted as $a_i$, thus invariant gauge tensors $T^{a_1,\dots,a_N}$ are needed to contract $N$ field building blocks $D^{w_i}\Psi_{i,a_i}$ in the operator to form a gauge singlet.
Therefore the operator involving $N$ fields at a certain dimension $d$ can be formally expressed as,
\beq\label{eq:OpeForm}
\mathcal{O}_N^{(d)}=T^{a_{1}, \ldots, a_{N}} \epsilon^{n} \tilde{\epsilon}^{\tilde{n}} \prod_{i=1}^{N} D^{w_i} \Psi_{i, a_{i}},
\eeq
where $n$ ($\tilde{n}$) number of $\epsilon$ ($\tilde{\epsilon}$) are used to contract all the undotted (dotted) spinor indices of the building blocks in Eq.~\eqref{eq:highestweight}. For a massless field $\Psi_i$, the helicity $h_i$ of the in-coming particle states that $\Psi_i$ annihilates indicates that $\Psi_i$ transforms as the $(-h_i,0)$ or $(0,h_i)$ irreducible representation of $SL(2,\mathbb{C})$ for $h_i<0$  and $h_i>0$ respectively. Under this assumptions, we have the following relations between $n$, $\tilde{n}$, $N$, $d$, $w_i$ and $h_i$:
\begin{eqnarray}
 & \tilde{n}+n \equiv r =\sum_{i} \left(\omega_{i}+|h_i| \right), \quad \tilde{n}-n = \sum_ih_i \equiv h, \quad d=\tilde{n}+n+N,\label{eq:spinor_index_matching_1}
\end{eqnarray}
which can be used to derive a number of inequalities to enumerate all the Lorentz classes for a given dimension $d$.

With the operator-amplitude correspondence introduced in Ref.~\cite{Li:2020gnx,Li:2020zfq,Li:2022tec}, one can easily relate the $T^{a_{1}, \ldots, a_{N}}$ and  $\epsilon$ ($\tilde{\epsilon}$) to the group factor and brackets of the local on-shell amplitudes generated by these operators in the spinor-helicity formalism. Therefore the powerful Young tableaux technique developed for the Lorentz and gauge structures of amplitudes can be used for constructing operator basis, which is the reason why we name it as the y-basis. The procedure for finding the complete and independent Lorentz and gauge structures with the Young tensor method is briefly summarized as follows.  

\bit 
\item The Lorentz sector is represented by a Young tensor component of the product group $SL(2,C) \times SU(N)$, introduced in Ref.~\cite{Henning:2019enq,Li:2020gnx,Li:2020zfq,Li:2020xlh,Li:2022tec}, where $N$ is the number of external particles in the on-shell amplitude.
All irreducible representations denoted by Young diagrams in the reduction vanish due to the momentum conservation and Schouten identity except the primary Young diagram
\begin{eqnarray}\label{eq:primary_YD}
\\ \nonumber
Y_{N,n,\tilde{n}} \quad = \quad \arraycolsep=0.2pt\def\arraystretch{1}
\rotatebox[]{90}{\text{$N-2$}} \left\{
\begin{array}{cccccc}
\yng(1,1) &\ \ldots{}&\ \yng(1,1)& \overmat{n}{\yng(1,1)&\ \ldots{}\  &\yng(1,1)} \\
\vdotswithin{}& & \vdotswithin{}&&&\\
\undermat{\tilde{n}}{\yng(1,1)\ &\ldots{}&\ \yng(1,1)} &&&
\end{array}
\right..
\\ \nonumber
\\
\nonumber 
\end{eqnarray}

Given a set of labels $\{1,\cdots,N\}$ corresponding to each particle in the local on-shell amplitude or equivalently to each fields in the operator type, the number of labels to be filled in the primary Young diagram is given by:
\beq\label{eq:numoflabels}
	\#i = \tilde{n} - 2h_i, \quad i=1,\cdots,N.
\eeq
All the Semi-Standard Young Tableau (SSYT) obtained by filling the labels $$\{\overbrace{1,\dots,1}^{\#1},\overbrace{2,\dots,2}^{\#2},\dots,\overbrace{N,\dots,N}^{\#N}  \} $$ in the primary Young diagram span the space of all amplitudes for the operator type, and each SSYT is a basis vector of this space and can be translated to a on-shell local amplitude and via amplitude-operator correspondence corresponds to a Lorentz structure of the operators, which we collectively denote as $\{{\cal B}^{(\rm y)}_i\}$.

\item As for the gauge factor $T$, an algorithm~\cite{Li:2020gnx,Li:2020zfq,Li:2020xlh,Li:2022tec} is proposed to find all the independent gauge Young tensor $\{T^{(\rm y)}_i\}$ expressed in terms of products of $M$th-rank Levi-Civita tensors of the $SU(M)$ group given that all the fields are expressed with fundamental indices only. The algorithm is to apply the generalized Littlewood-Richardson rules to construct the singlet Young tableaux from the set of Young tableaux corresponding to each field.  

\eit

The direct product of the Lorentz Young tensor $\{{\cal B}^{(\rm y)}_i\}$ and gauge Young tensor $\{T^{(\rm y)}_i\}$ results in the complete and independent operator basis if no repeated fields are encountered, or equivalently if all the particles in corresponding amplitude are distinguishable. These are referred to as \emph{flavor-blind} operators as flavor indices are treated as labels that distinguish the fields rather than indices that may take equal values. 
One necessary comment is that the y-basis operators may not be monomials if converted to a form with Lorentz indices rather than spinor indices in the Lorentz sector, or adjoint (or higher representation) indices rather than fundamental indices in the gauge sector. However, since we provide a subroutine to find coordinates of any operators, including the monomial ones, under the complete flavor-blind y-basis, it is easy to select an independent and complete set of monomial operators ${\cal O}^{(\rm m)}$, which we call the m-basis.

In the presence of repeated fields, the flavor-blind operators get more redundancies from the permutation symmetries among them, which are referred to as flavor relations when they have flavor indices like the SM fermions. When these redundancies are considered, the operator is said to be \emph{flavor-specified}.
The complete basis of flavor-specified operators can be most easily obtained by organizing the flavor-blind operators into irreducible representations of the permutation group among the repeated fields, which we call the p-basis. The corresponding Young diagrams for the irreducible representations become Young tableau when filled with the flavor indices\footnote{When there is no flavor or $n_f=1$, like the repeated bosons in the SM, an auxiliary flavor index is understood in the flavor-blind operator so that they can be distinguished. At the flavor-specified level, these auxiliary flavor indices can take only one value when the flavor Young tensors are counted. }, which constitute irreducible tensor representations, namely the Young Tensor, of the corresponding (auxiliary) flavor $SU(n_f)$ group. As a result, the SSYTs of the Young Tensor represent the independent flavor-specified operators. By the observation from the following 
\begin{eqnarray}
\underbrace{\pi\circ {\cal O}^{\{f_{k},...\}}}_{\rm permute\ flavor} &=& \underbrace{\left(\pi\circ T_{{\rm G_1}}^{\{g_k,...\}}\right)\left(\pi\circ T_{{\rm G_2}}^{\{h_k,...\}}\right)\cdots}_{\rm permute\ gauge}\underbrace{\left(\pi\circ{\cal B}^{\{f_k,...\}}_{\{g_{k},...\},\{h_{k},...\}}\right)}_{\rm permute\ Lorentz},
\label{eq:tperm}
\end{eqnarray}
we can construct symmetrizers of the flavor indices by applying the permutations to the gauge and Lorentz indices of the building blocks. A matrix representation of the symmetrizer can be obtained by using the subroutine of finding coordinates of the permuted operators under the y-basis or m-basis, while the p-basis can be selected as the independent rows. The operator obtained in this way is thus in the following form
\begin{equation}
    \mathcal{O}^{(\rm p)}_{prst} = \mc{Y}[{\tiny\young(prs,t)}]\circ\mathcal{O}^{(\rm m)}_{prst}
\end{equation}
where $\mc{Y}[{\tiny\young(prs,t)}]$ is the Young symmetrizer for the flavor symmetry ${\tiny\yng(3,1)}$, and the flavor indices taking values that satisfy the SSYT condition serve as an independent operator basis for this Young Tensor. Note that in practice the Young symmetrizer may not be actually applied to the operator rendering a polynomial, but can be understood as acting on the Wilson coefficient tensor $C^{prst}$ so that flavor relations are obtained.

The above algorithm not only provides a systematical way to obtain the complete and independent operator on-shell basis of the SMEFT at any mass dimension, but can also be applied to generic EFTs with arbitrary scalar, fermion and gauge extensions, such as the left-right symmetric model, grand unification, etc. 
Various bases of operators for different purposes are defined, and the conversions among them are made easy using the reduction algorithm implemented in the Mathematica package ABC4EFT~\footnote{https://abc4eft.hepforge.org}.

\section{All-orders results in the \texorpdfstring{$1/\Lambda$}{1/Lambda} expansion} \label{sec:allorders}

In this white paper, we review the construction of the ‘geometric’ basis~\cite{Helset:2020yio, Hays:2020scx} and how it can be used to study $1 \to 2$ process at $\mathcal O(1/\Lambda^4)$. These processes serve as a laboratory to explore truncation error from higher orders (in $1/\Lambda$) terms, a necessary ingredient in SMEFT global fits.

When considering higher dimensional operators in SMEFT, operators that contribute (in the broken phase) to 2- and 3-particle vertices are particularly important because they feed into how SM fields are defined and how the parameters of the SM are related to experimental observables. 

While the number of operators at $d > 6$ grows rapidly, the number of operators that contribute to 2- and 3-point vertices is small and is approximately constant at each mass dimension. This may seem counter intuitive, as one may think it's always possible to staple on more derivatives or powers of the Higgs field onto an operator of dimension $d$ to generate an even higher dimensional term. However, the kinematics for 2- and 3-point vertices is trivial, meaning all dot products of momenta are related to masses of the particles involved, so e.g. adding two more derivatives to an operator with dimension $d$ does not result in a new operator structure with dimension $d+2$. Second, with only two or three fields around, there are limited electroweak structures possible, and therefore limited ways to dress up the operator with Higgses. 

As a consequence, one can determine the operators that contribute to 2- and 3-point vertices to {\em all orders} in $v_T/\Lambda$\footnote{Here $v_T$ is the minimum of the full Higgs potential including higher order terms, and is distinct from the SM Lagrangian parameter $v_0$. It is possible to express the former in terms of the latter, though this is only necessary in processes where multiple Higgses are produced, as the parameter that enters the $W/Z$ masses -- and is therefore linked to $G_F$ -- is $v_T$.}. For example, the operators that feed into the kinetic term for electroweak gauge bosons are limited to
\begin{align}
 Q_{HB}^{(6+2n)} &= (H^\dagger H)^{n+1} B^{\mu \nu} \, B_{\mu \nu},  \\
Q_{HW}^{(6+ 2n)} &=  (H^\dagger H)^{n+1} W_a^{\mu \nu}\, W^a_{\mu \nu}, \\
Q_{HWB}^{(6+ 2n)} &=  (H^\dagger H)^n (H^\dagger \sigma^a H) W_a^{\mu \nu}\, B_{\mu \nu}, \\
Q_{HW,2}^{(8+ 2n)} &=  (H^\dagger H)^{n} (H^\dagger \sigma^a H) (H^\dagger \sigma^b H) W_a^{\mu \nu}\, W_{b,\mu \nu},
\end{align}
Re-expressing the operators in terms of the four real degrees of freedom $\phi_I$ in the Higgs and combining them with the SM terms, we can lump all of the SMEFT effects into a `metric' $g_{AB}(\phi)W^A_{\mu \nu} W^{B,\mu \nu}$. Explicitly
\begin{align}
\label{eq:gabmetric}
g_{AB}(\phi_I) &= \left[1 
-4\sum_{n=0}^\infty \left(C_{HW}^{(6+ 2n)}(1- \delta_{A4}) + C_{HB}^{(6+ 2n)} \delta_{A4}\right) \left(\frac{\phi^2}{2}\right)^{n+1} \right]\delta_{AB} \nonumber \\
&+  \sum_{n=0}^\infty C_{HW,2}^{(8+ 2n)} \, \left(\frac{\phi^2}{2}\right)^{n} \left(\phi_I \Gamma_{A,J}^I \phi^J\right) \,  \left(\phi_L \Gamma_{B,K}^L \phi^K\right) (1- \delta_{A4})(1- \delta_{B4}) \nonumber \\
&+\left[\sum_{n=0}^\infty C_{HWB}^{(6+ 2n)} \left(\frac{\phi^2}{2}\right)^n \right] (\phi_I \Gamma_{A,J}^I \phi^J) \, (1-\delta_{A4})\delta_{B4},
\end{align}
where $\Gamma^A$ are the $SU(2)$ generators as 4 by 4 matrices (see  for definition).

Carrying out this process for all other possible 2- and 3- point vertices using SM fields, we arrive at a set of metrics:
\begin{align*}
& h_{IJ}(\phi) (D_\mu \phi)^I (D_\mu \phi)^J,  \quad g_{AB}(\phi) W^A_{\mu \nu} W^{B,\mu \nu},  \quad
k_{IJ}^A(\phi) (D_\mu \phi)^I (D_\nu \phi)^J \, W_{A}^{\mu \nu},  \quad f_{ABC}(\phi)  W^{A}_{\mu \nu} W^{B, \nu \rho} W^{C, \rho \mu}, \\
& \quad\quad\quad\quad\quad\quad \quad\quad\quad  Y(\phi) \bar{\psi}_1 \psi_2,  \quad L_{I,A}(\phi) \bar{\psi}_1 \gamma^\mu \tau_A \psi_2 (D_\mu \phi)^I,  \quad d_A(\phi) \bar{\psi}_1 \sigma^{\mu \nu} \psi_2 W^A_{\mu \nu},
\end{align*}
(plus analogous forms for gluons and with dual field strengths). Each possesses an expansion similar to those found in ~\cite{Hays:2018zze} in terms of a small number of easily identifiable operators at each mass dimension.

Working with the metric forms, canonically normalizing the gauge bosons and changing to the mass eigenstate basis, we can identify the couplings of matter to $W/Z$ in the presence of higher dimensional operators.
\begin{align}
\langle \mathcal{Z} | \bar{\psi}_{\substack{p}} \psi_{\substack{r}}\rangle &=
\frac{\bar{g}_Z}{2} \, \bar{\psi}_{\substack{p}} \, \slashed{\epsilon}_{\mathcal{Z}} \, \left[(2 s_{\theta_Z}^2  Q_\psi  - \sigma_3)\delta_{pr}
	+\sigma_3\bar{v}_T \langle L_{3,3}^{\psi,pr}\rangle+ \bar{v}_T \langle L_{3,4}^{\psi,pr} \rangle
 \right] \, \psi_{\substack{r}}, \\
\langle \mathcal{W}_{\pm} | \bar{\psi}_{\substack{p}} \psi_{\substack{r}} \rangle &= - \frac{\bar{g}_2  }{\sqrt{2}}
\bar{\psi}_{\substack{p}}  (\slashed{\epsilon}_{\mathcal{W}^{\pm}}) \, T^{\pm}
\left[\delta_{pr}-\bar{v}_T \langle L^{\psi,pr}_{1,1}\rangle   \pm  i \bar{v}_T
\langle L^{\psi,pr}_{1,2} \rangle \right]
\, \psi_{\substack{r}}.
\end{align}
Here the $\langle \rangle$ indicates taking the vacuum expectation value and $p,r$ are flavor labels. The couplings and mixing angles are now defined to all orders in terms of the metrics
\begin{align}
\bar{g}_2 &= g_2 \, \sqrt{g}^{11} = g_2 \, \sqrt{g}^{22}, \nonumber \\
\bar{g}_Z &= \frac{g_2}{c_{\theta_Z}^2}\left(c_{\bar{\theta}} \sqrt{g}^{33} - s_{\bar{\theta}} \sqrt{g}^{34} \right) = \frac{g_1}{s_{\theta_Z}^2}\left(s_{\bar{\theta}} \sqrt{g}^{44} - c_{\bar{\theta}} \sqrt{g}^{34} \right), \nonumber \\
\bar{e} &= g_2\left(s_{\bar{\theta}} \sqrt{g}^{33} + c_{\bar{\theta}} \sqrt{g}^{34}  \right) = g_1\left(c_{\bar{\theta}} \sqrt{g}^{44} + s_{\bar{\theta}} \sqrt{g}^{34}  \right), \nonumber \\
s_{\theta_Z}^2 &= \frac{g_1 (\sqrt{g}^{44} s_{\bar{\theta}} - \sqrt{g}^{34} c_{\bar{\theta}})}{g_2 (\sqrt{g}^{33} c_{\bar{\theta}} - \sqrt{g}^{34} s_{\bar{\theta}})+ g_1 (\sqrt{g}^{44} s_{\bar{\theta}} - \sqrt{g}^{34} c_{\bar{\theta}})}, \nonumber \\
s_{\bar{\theta}}^2 &= \frac{(g_1 \sqrt{g}^{44} - g_2 \sqrt{g}^{34})^2}{g_1^2 [(\sqrt{g}^{34})^2+ (\sqrt{g}^{44})^2]+ g_2^2 [(\sqrt{g}^{33})^2+ (\sqrt{g}^{34})^2]
- 2 g_1 g_2 \sqrt{g}^{34} (\sqrt{g}^{33}+ \sqrt{g}^{44})}.
\end{align}
where we have used the notation $\sqrt{g^{11}} = \langle \sqrt{g^{-1}}\rangle_{11}$, etc.

Other trilinear couplings, such as the coupling between the Higgs and two photons, can be obtained by extracting the linear $h$ term from the $g_{AB}$ metric,
\begin{align}
\langle h| \mathcal{A}(p_1) \mathcal{A}(p_2) \rangle = -\langle h A^{\mu\nu} A_{\mu \nu} \rangle \frac{\sqrt{h}^{44}}{4} \left[
	\langle \frac{\delta g_{33}(\phi)}{\delta \phi_4}\rangle \frac{\overline{e}^2}{g_2^2}  +
 2\langle \frac{\delta g_{34}(\phi)}{\delta \phi_4}\rangle \frac{\overline{e}^2}{g_1 g_2}  +
  \langle \frac{\delta g_{44}(\phi)}{\delta \phi_4}\rangle \frac{\overline{e}^2}{g_1^2}
	\right], \nonumber
\end{align}

The above compact formulae allow us to study a variety of phenomenologically interesting processes at the next EFT order $\mathcal O(1/\Lambda^4)$ with relative ease and without an explosion in the number of coefficients. For example, $h \to \gamma\gamma$ production at $\mathcal O(1/\Lambda^4)$ is~\cite{Hays:2020scx}:
\bea\label{correctsquarehgamgam}
|\langle h| \gamma \gamma\rangle|^2_{{\rm{to}} \, {\cal{O}}(v^4/\Lambda^4)} &=&  \bar{v}_T^2\bigg|\mathcal{A}_{\rm SM}^{h\gamma\gamma}  \bigg|^2 + 2\bar{v}_T \,{\rm Re}(\mathcal{A}_{\rm SM}^{h\gamma\gamma})
(1+\langle \sqrt{h}^{44}\rangle_{{\cal{O}}(v^2/\Lambda^2)}) \, \langle h|\gamma \gamma\rangle_{\mathcal{L}^{(6)}} \\
&+&(1+ 4 \,\bar{v}_T \, {\rm Re}(\mathcal{A}_{\rm SM}^{h\gamma\gamma})) \, \langle h|\gamma \gamma\rangle_{\mathcal{L}^{(6)}}^2
+ 4\bar{v}_T \,{\rm Re}(\mathcal{A}_{\rm SM}^{h\gamma\gamma})
\left.(\langle h|\gamma \gamma\rangle_{\mathcal{L}^{(8)}})\right. . \nonumber
\eea
where $\mathcal{A}_{\rm SM}^{h\gamma\gamma}$ is the SM loop level contribution and 
\begin{align}
\langle \sqrt{h}^{44}\rangle_{{\cal{O}}(v^2/\Lambda^2)}
	&= \tilde C_{H\Box}^{(6)} - \frac{1}{4}\tilde C_{HD}^{(6)},  \nonumber \\  
	\langle h|\gamma \gamma\rangle_{\mathcal{L}^{(6)}} &= \left[\frac{g_2^2 \, \tilde C_{HB}^{(6)}
  + g_1^2 \, \tilde C_{HW}^{(6)} - g_1 \, g_2 \, \tilde C_{HWB}^{(6)}}{({g}^{\rm SM}_Z)^2} \right], \nonumber \\
  	\langle h|\gamma \gamma\rangle_{\mathcal{L}^{(8)}} &= \left[\frac{g_2^2 \, \tilde C_{HB}^{(8)}
  + g_1^2 \, (\tilde C_{HW}^{(8)} + \tilde C_{HW,2}^{(8)}) - g_1 \, g_2 \, \tilde C_{HWB}^{(8)}}{2({g}^{\rm SM}_Z)^2} \right].
\end{align}
Here the tilded coefficients include powers of $v^2_T/\Lambda^2$, i.e. $\tilde C^{(6)}_i = C^{(6)}_i v^2_T/\Lambda^2, \tilde C^{(8)}_i = C^{(8)}_i v^4_T/\Lambda^4$. From this result, we see that only a handful, $\mathcal O(8)$ coefficients are involved and we can identify the situations where naive power counting (dimension 8 less important that dimension 6) breaks down, such as when dimension 6 operators are generated at loop level while dimension 8 operators are generated at tree level. See Ref.~\cite{Hays:2020scx, Corbett:2021eux} for explicit examples of this tree/loop generation and Ref.~\cite{Arzt:1994gp, Arzt:1994gp,Craig:2019wmo} for a general classification of operators by the loop order they are generated at in weakly coupled UV theories.

Further examples, such as $h \to Z\gamma, h \to ZZ^*, Z \to \bar{\psi} \psi$ can be found in Ref.~\cite{Hays:2020scx}, and the $\mathcal O(1/\Lambda^4)$ results for $h \to \gamma\gamma$, $h \to g g$, and $gg \to h$ have been combined with one loop  perturbative QCD corrections in Ref.~\cite{Corbett:2021cil, Martin:2021vwf}.  The main goal of this line of study is to extend the list of processes known to $\mathcal O(1/\Lambda^4)$, with the hope that a broader index of exact results will improve how truncation uncertainties in EFT analyses are estimated.

\section{Renormalization group running and implications for positivity}
\label{sec:dim8rges}
The renormalization group evolution (RGE) and mixing of all dimension-8 SMEFT operators is not completely known yet, but substantial progress has been made in this regard in recent years.

There exist two co-leading contributions to the RGEs of dimension-8 operators, namely those arising from loops involving two dimension-6 interactions and those with only one dimension-8 term. (While the dimension-6 RGEs have also two contributions, that with two dimension-5 terms~\cite{Davidson:2018zuo,Chala:2021juk} is most probably negligible due to the large lepton-number violation scale.)  In the usual approach to renormalization, based on the computation of 1-particle-irreducible (1PI) Feynman diagrams off-shell, these loops generate divergences that can not be absorbed by physical interactions alone (as those described in sections~\ref{sec:opcount} and \ref{sec:dim8}), but they require including operators that only later can be removed from the action via field redefinitions. A basis of physically-independent interactions can be extended to a basis of independent Green's functions~\cite{Jiang:2018pbd}, the elements of which can not be related among themselves by integration by parts, nor by Fierz, Bianchi or algebraic identities. Knowing one such Green's basis for the SMEFT can ease significantly the process of renormalization, as well as the matching of UV models onto the SMEFT.
 
Thus, a SMEFT dimension-6 Green's basis was worked out in Ref.~\cite{Gherardi:2020det}. It consists of 81 independent interactions, that extend the 59 physical operators. The dimension-8 counterpart involves a much larger number of terms, and so far only a basis of bosonic interactions is known~\cite{Chala:2021cgt}; it involves 86 new operators. The way to obtain this basis relies on the observation that, given $N$ effective operators $\lbrace\mc{O}_i\rbrace_{i=1,...N}$ with Wilson coefficients $c_i$, their contribution to an off-shell 1PI amplitude $\mathcal{A}(a\to b)$ reads:
\begin{equation}
 \mathcal{A}(a\to b) = \sum_{\alpha\in I} c_i f^i_\alpha(\vec{g})\kappa_\alpha\,,
\end{equation}
where $I$ denotes a collection of indices, $f^i_\alpha$ is a matrix which is only function of the SM couplings $\vec{g}=(g_1,g_2,g_3,\lambda)$ and $\lbrace \kappa_\alpha\rbrace_{\alpha\in I}$ are \textit{independent} kinematic invariants. If $f^i_\alpha$ has rank $N$, then the operators are off-shell independent. 
%

On the basis of these results, the first systematic computation of the SMEFT RGEs to order $v^4/\Lambda^4$ was initiated in Ref.~\cite{Chala:2021pll}. This includes the renormalization of both relevant and marginal (to dimension-8) interactions as triggered by loops involving two insertions of dimension-6 operators that can arise at tree-level in weakly-coupled UV completions of the SMEFT; e.g. $\mathcal{O}_{H\square} = (H^\dagger H)\square (H^\dagger H)$. (The effects of loops involving other type of operators, such as $\mc{O}_{HW} = W_{\mu\nu}^I W^{I\mu\nu} (H^\dagger H)$, which can only arise at loop level~\cite{Craig:2019wmo}, are formally two-loop corrections.) Among the main results that can be highlighted from this computation, we find:

\begin{enumerate}
 \item No loop-generated operator is renormalized by pairs of tree-level dimension-6 interactions. This result extends the previous findings at dimension-6~\cite{Elias-Miro:2013gya}.
 \item Certain anomalous dimensions vanish due to cancellations between different operators that occur only on-shell. These zeros are not yet explained by non-renormalization theorems such as those in Refs.~\cite{Cheung:2015aba}. Also, one of these zeros implies that the Peskin-Takeuich parameters $S$ and $U$~\cite{Peskin:1990zt} are not renormalized by tree-level dimension-6 terms (to order $v^4/\Lambda^4$). 
 %
 \item The indirect constraints on some Wilson coefficients derived from their effects on electroweak observables are competitive with direct bounds from facilities such as the LHC. 
 
 %
\end{enumerate}
%

The RGEs ensuing from loops involving one insertion of dimension-8 operators have been considered in Ref.~\cite{AccettulliHuber:2021uoa}. The authors use an on-shell approach to the SMEFT (see next section); restricting the calculation to linear order in the Higgs quartic coupling $\lambda$ and to quadratic order in the gauge couplings $g$, and neglecting corrections, proportional to the Higgs mass $m_H$, to the RGEs of lower-dimensional operators. These missing contributions in Ref.~\cite{AccettulliHuber:2021uoa} have been computed, within the SMEFT bosonic sector, in Ref.~\cite{Bakshi:2022abc}. Among other interesting aspects of this result, one can highlight the presence of loop-induced operators (of both dimension-8 and dimension-6) that get renormalized by tree-level interactions. For example:
\begin{align}
 16\pi^2 \mu \frac{d}{d\mu}c_{W^2 H^2 D^2}^{(1)} &= \frac{g_2^2}{6} (2 c_{H^4}^{(1)}+3 c_{H^4}^{(2)}+c_{H^4}^{(3)})\,,\\
 16\pi^2 \mu\frac{d}{d\mu} c_{H WB} &= m_H^2 \left[\frac{g_1 g_2}{2} (c_{H^4}^{(1)}-2c_{H^4}^{(2)}+c_{H^4}^{(3)})+8 c_{WB H^4}^{(1)}+\cdots\right]\,,
\end{align}
where $\mc{O}_{W^2 H^2 D^2}^{(1)} = (D^\mu H^\dagger D^\nu H) W_{\mu\rho}^I W^{I\,\rho}_\nu$ and $\mc{O}_{WBH} = W_{\mu\nu}^I B^{\mu\nu}(H^\dagger\sigma^I H)$, whereas $\mc{O}_{WBH^4}^{(1)}=(H^\dagger H)(H^\dagger\sigma^I H)W_{\mu\nu}^I B^{\mu\nu}$, $\mc{O}_{H^4}^{(1)} = (D_\mu H^\dagger D_\nu H)(D^\nu H^\dagger D^\mu H)$, 
$\mc{O}_{H^4}^{(2)} = (D_\mu H^\dagger D_\nu H)(D^\mu H^\dagger D^\nu H)$ and $\mc{O}_{H^4}^{(3)} = (D^\mu H^\dagger D_\mu H)(D^\nu H^\dagger D_\nu H)$.

The knowledge of the dimension-8 RGEs has also important implications for the so-called \textit{positivity bounds}. These bounds are restrictions on the sign of (combinations of) Wilson coefficients that ensue from the very basic principles of analiticity and unitarity of the S-matrix~\cite{Adams:2006sv}. These constraints are of paramount importance not only because experimental evidence of the violation of positivity in the data would imply the potential breakdown of either relativity or quantum mechanics, but more realistically because these restrictions can modify substantially the ``priors'' in experimental fits aiming at constraining the SMEFT parameter space~\cite{Zhang:2018shp,Bi:2019phv}.

Bounds of this type have been derived both at tree level (assuming that the only relevant singularities of the S-matrix are single poles) as well as at one-loop (i.e. in the presence of branch cuts). It has been shown~\cite{Chala:2021wpj} though, that these bounds are in general not scale-invariant. In other words, even if they hold at some scale $\mu=\Lambda$, they can be broken by the RGEs at scales $\mu\ll\Lambda$. For example, the following inequalities hold at tree level~\cite{Bi:2019phv,Remmen:2019cyz}:
\begin{align}\label{eq:positivity1}
 c_{H^4}^{(2)}&> 0\,,\\
 c_{H^4}^{(1)} + c_{H^4}^{(2)} &> 0\,,\\
 c_{H^4}^{(1)} + c_{H^4}^{(2)} + c_{H^4}^{(3)} &> 0\,.\label{eq:positivity2}
\end{align}
However, the RGEs of these Wilson coefficients can take them out of the positivity region. For example, assuming $c_{H^4}^{(2)}(\Lambda)=0$, we obtain~\cite{AccettulliHuber:2021uoa,Chala:2021wpj,Bakshi:2022abc}:
\begin{equation}
 c_{H^4}^{(2)}(\mu) = \frac{1}{96\pi^2}\left[28 c_{H^4}^{(1)}(\Lambda) + 
 15 c_{H^4}^{(3)}(\Lambda)\right] g_2^2\log{\frac{\mu}{\Lambda}} + \mc{O}(g_1^2,\lambda)\,,
\end{equation}
which can be clearly negative even for values of the Wilson coefficients satisfying Eqs.~\eqref{eq:positivity1}--\eqref{eq:positivity2} at $\mu=\Lambda$ at which they arise at tree level.

The story is significantly different for other anomalous-gauge-quartic-coupling operators. For example, Ref.~\cite{Bi:2019phv} shows that:
\begin{equation}
 -2 f_{M,1}+f_{M,7} > 0\,;
\end{equation}
see the reference above for the definition of these Wilson coefficients. In the basis of Ref.~\cite{Murphy:2020rsh}, this expression reads $c_{W^2 H^2 D^2}^{(1)}<0$~\cite{Bakshi:2022abc}. It holds at all scales within one-loop accuracy, given that this operator is not renormalized by pairs of dimension-6 interactions~\cite{Chala:2021pll} and because~\cite{Bakshi:2022abc}:
\begin{align}
 c_{W^2H^2 D^2}^{(1)}(\mu) = c_{W^2H^2 D^2}^{(1)}(\Lambda)-\frac{g_2^2}{96\pi^2} \left[2 c_{H^4}^{(1)}(\Lambda)+3 c_{H^4}^{(2)}(\Lambda)+c_{H^4}^{(3)}(\Lambda)\right]\log{\frac{\Lambda}{\mu}}\,,
\end{align}
where the first term vanishes (because $c_{W^2H^2 D^2}^{(1)}$ arises only at one loop~\cite{Craig:2019wmo}) and the second term is negative because Eqs.~\eqref{eq:positivity1}--\eqref{eq:positivity2} do hold at tree level. The same is valid for all other inequalities obtained in Ref.~\cite{Bi:2019phv} involving $f_{M,i}$, $i=1,...,5,7$~\cite{Bakshi:2022abc}. This result demonstrates that, contrary to the constraints on the $c_{H^4}^{(i)}$ couplings, those on $f_{M,i}$ can be consistently enforced as Bayesian ``priors'' in experimental fits aiming at measuring these parameters.

\section{On-shell approach to the SMEFT}

On-shell amplitude techniques provide an alternative approach to the SMEFT, notably avoiding the gauge and field-redefinition redundancies inherent in the Lagrangian treatment.
This is particularly useful to explore patterns and properties arising from scattering amplitudes in the presence of higher-dimensional operators.

For example, the structure of the anomalous dimension matrix up to higher loop orders and higher operator dimensions~\cite{Cheung:2015aba, Bern:2019wie, Jiang:2020sdh, EliasMiro:2020tdv, Baratella:2020lzz, Jiang:2020mhe, Bern:2020ikv, Baratella:2020dvw, Shu:2021qlr, AccettulliHuber:2021uoa, EliasMiro:2021jgu}, non-interference theorems~\cite{Jiang:2020sdh, Azatov:2016sqh}, symmetry selection rules~\cite{Jiang:2020sdh} as well as sum rules~\cite{Gu:2020thj} are made manifest in this approach.
On the other hand, the direct construction of the non-factorizable on-shell amplitudes for massless particles~\cite{Shadmi:2018xan, Ma:2019gtx, Falkowski:2019zdo, Durieux:2019siw, Li:2020zfq, Li:2020gnx, Li:2020xlh, Li:2022tec}  efficiently substitutes the enumeration of operators.
The approach adopted in~\cite{Li:2020zfq, Li:2020gnx, Li:2020xlh, Li:2022tec} is detailed in \autoref{sec:dim8_YT}.

The recent development of a little-group-covariant formalism~\cite{Arkani-Hamed:2017jhn} allowed to apply on-shell techniques to massive particles of arbitrary spin.
The renormalizable SM amplitudes were studied in~\cite{Christensen:2018zcq, Christensen:2019mch, Bachu:2019ehv} and the map between the massive three-point on-shell amplitudes to dimension-6 operators in the Warsaw basis were presented in~\cite{Aoude:2019tzn,Durieux:2019eor}.
A further step was taken in~\cite{Durieux:2019eor}, where the electroweak symmetry is not built-in but can be recovered by imposing perturbative unitarity.
Moreover, a systematic construction of three and four-point non-factorizable amplitudes was presented in~\cite{Durieux:2020gip}.
Systematic algorithms for the construction of independent massive amplitudes were further in~\cite{Dong:2021yak, DeAngelis:2022qco, Dong:2022mcv}, while their derivation from the Higgsing of massless amplitudes was studied in~\cite{Balkin:2021dko}.
This approach also yields all-order results in $v/\Lambda$, whose powers are all absorbed in constant amplitude coefficients.
Tree-level recursion relations for massive amplitudes have been investigated too~\cite{Franken:2019wqr, Falkowski:2020aso}.

The development of this alternative approach to the SMEFT provides new insight and allows for more efficient computations.

\section{Novel phenomenological consequences at dimension-8}
\label{sec:dim8pheno}

The naive expectation is that deviations from the SM induced by
dimension-8 operators are subdominant to dimension-6 deviations and
can be safely ignored.  While this is sometimes the case, the
increasing precision of LHC data is beginning to require the inclusion
of even such subleading effects in global fits.  There are additionally
interesting cases where the dimension-8 terms are sometimes the leading contributions
to observables due to symmetry considerations or the structure of the
corresponding SM amplitudes.  In such cases it is important to quantify their effects in order to guide
experimental searches.  Such probes may also serve as smoking-gun
signatures of dimension-8 extensions of the SM.

\subsection{Impact of dimension-8 operators on Drell-Yan angular distributions}\label{angular}
We discuss in this white paper both the impact of dimension-8 operators in SMEFT fits, and examples where dimension-8 effects
give qualitatively different results than dimension-6, using the Drell-Yan process as an example. Drell-Yan is one of the best measured and calculated processes at the LHC, with the residual uncertainties from experiment and uncalculated theory approaching the percent level. It therefore serves as a test case for future high-luminosity runs of the LHC where additional processes may reach a similar precision benchmark. In this section we consider the angular distribution of leptons in the Drell-Yan process following the recent study of Ref.~\cite{Alioli:2020kez}, while in the next section we investigate the impact of dimension-8 terms on global fits to SMEFT parameters.

The standard theoretical formalism was developed in seminal work several decades ago~\cite{Collins:1977iv}. The form of the angular distribution to all orders in the strong coupling constant follows from the spin-1 nature of the photon and $Z$-boson which mediate the interaction:
\begin{eqnarray}
\frac{d \sigma}{dm_{ll}^2 dy d\Omega_l} &=& \frac{3}{16\pi} \frac{d
  \sigma}{dm_{ll}^2 dy}
\left\{(1+c_{\theta}^2)+\frac{A_0}{2}(1-3c_{\theta}^2) \right. \nonumber \\
 && \left. +A_1
s_{2\theta}c_{\phi}+\frac{A_2}{2}s^2_{\theta}c_{2\phi} +A_3
                                         s_{\theta}c_{\phi}+A_4 c_{\theta}\right. \nonumber \\
 && \left.  +A_5 s^2_{\theta}s_{2\phi} +A_6 s_{2\theta}s_{\phi} +A_7
    s_{\theta}s_{\phi}
     \right\}.
     \label{eq:oldCSexp}
\end{eqnarray}
Here, $m_{ll}$ is the invariant mass of the lepton system, $y$ is the
rapidity of the $Z$-boson that produces the lepton pair, and
$\Omega_l$ is the solid angle of a final-state lepton. The lepton
angles are typically defined in the Collins-Soper
frame~\cite{Collins:1977iv} and we have used the notation $s_{\alpha}$
and $c_{\alpha}$ to represent their sine and cosine, respectively.  In
the SM, the leptons are produced by an $s$-channel spin-one current,
so in the squared amplitude spherical harmonics up to $l=2$ are
allowed. 

In the SMEFT, however, there is a class of two-derivative dimension-8 operators that populate the $l=2$ partial wave at the amplitude level,
allowing for $l=3$ spherical harmonics in the angular expansion when
interfered with the SM amplitude. Dimension-6 operators cannot
generate $l=2$ partial waves at the amplitude level, making their
appearance a hallmark of the dimension-8 SMEFT. Following Ref.~\cite{Alioli:2020kez} we express these operators as
   \begin{eqnarray}
     {\cal O}_{8,lq\partial 3} &=& (\bar{l} \gamma_{\mu} \overleftrightarrow{D}_{\nu} l)
                                 (\bar{q} \gamma^{\mu} \overleftrightarrow{D}^{\nu} q)
                                    ,\nonumber \\
      {\cal O}_{8,lq\partial 4} &=& (\bar{l}\tau^I \gamma_{\mu} \overleftrightarrow{D}_{\nu} l)
                                 (\bar{q}\tau^I \gamma^{\mu}
                                    \overleftrightarrow{D}^{\nu} q), \nonumber \\
 {\cal O}_{8,ed\partial 2} &=& ( \bar{e}  \gamma_{\mu} \overleftrightarrow{D}_{\nu} e)( \bar{d}
 \gamma^{\mu} \overleftrightarrow{D}^{\nu} d),  \nonumber
   \\
 {\cal O}_{8,eu\partial 2} &=& ( \bar{e}  \gamma_{\mu} \overleftrightarrow{D}_{\nu} e)( \bar{u}
 \gamma^{\mu} \overleftrightarrow{D}^{\nu} u), \nonumber
     \\
{\cal O}_{8,ld\partial 2} &=&( \bar{l}  \gamma_{\mu} \overleftrightarrow{D}_{\nu} l)( \bar{d}
 \gamma^{\mu} \overleftrightarrow{D}^{\nu} d),  \nonumber
   \\
{\cal O}_{8,lu\partial 2} &=&( \bar{l}  \gamma_{\mu} \overleftrightarrow{D}_{\nu} l)( \bar{u}
 \gamma^{\mu} \overleftrightarrow{D}^{\nu} u),  \nonumber
   \\
{\cal O}_{8,qe\partial 2} &=& ( \bar{e}  \gamma_{\mu} \overleftrightarrow{D}_{\nu} e)( \bar{q}
                               \gamma^{\mu} \overleftrightarrow{D}^{\nu} q).
                               \label{eq:dim8d2ops}
   \end{eqnarray}
Here, $q$ and $l$ represent left-handed quark and lepton doublets respectively, $u,d,e$ correspond to right-handed singlets, and $\overleftrightarrow{D}_{\mu}= \overrightarrow{D}_\mu - \overleftarrow{D}_\mu$. In order to demonstrate the effect of these operators we calculate the contribution of ${\cal O}_{8,lq \partial 3}$ to the up-quark partonic matrix-element squared:
\begin{eqnarray}
\Delta |{\cal M}_{u\bar{u}}|^2 &=& -\frac{C_{8,lq\partial 3}}{\Lambda^4}
\,{\hat c}_{\theta}(1+\hat c_{\theta})^2 \frac{\hat{s}^2}{6} \times \nonumber \\ & & \left[e^2 Q_u
  Q_e  +\frac{g^2 g_L^u g_L^e \hat{s}}{c_W^2(\hat{s}-M_Z^2)} \right].
\label{eq:CLLd3amp}
\end{eqnarray}
Here, $\hat{s}$ denotes the usual partonic Mandelstam invariant
$\hat{s} = (p_1+p_2)^2$, $g$ is the $SU(2)$ coupling constant, $c_{W}$
is the cosine of the weak mixing angle, $e$ is the $U(1)_{EM}$
coupling constant, $Q_i$ is the charge of fermion $i$, $g_L^i$ are
the left-handed couplings to the $Z$-boson following the notation of Ref.~\cite{Denner:1991kt}.
$C_{8,lq\partial 3}$ is the Wilson coefficient associated with the
operator under consideration, and $\hat c_{\theta}$ is the angle between
the beam direction and the outgoing lepton direction. This contribution to the differential cross section contains a
$c_{\theta}^3$ dependence that cannot be described by
Eq.~(\ref{eq:oldCSexp}).  The reason for this was given in the
previous section when discussing the operators of
Eq.~(\ref{eq:dim8d2ops}): the traditional formulation of the lepton angular distribution is produced in the
$s$-channel by a spin-one current, which is not the case for $ {\cal
  O}_{8,lq\partial 3}$.  Only the seven dimension-8 operators identified above lead to an angular dependence not already described by
Eq.~(\ref{eq:oldCSexp}).

In order to account for this new signature of dimension-8 effects we extend the parameterization of Eq.~(\ref{eq:oldCSexp}) to
the following:
\begin{eqnarray}
\frac{d \sigma}{dm_{ll}^2 dy d\Omega_l} &=& \frac{3}{16\pi} \frac{d
  \sigma}{dm_{ll}^2 dy}
\left\{(1+c_{\theta}^2)+\frac{A_0}{2}(1-3c_{\theta}^2) \right. \nonumber \\
 && \left. +A_1
s_{2\theta}c_{\phi}+\frac{A_2}{2}s^2_{\theta}c_{2\phi} +A_3
                                         s_{\theta}c_{\phi}+A_4 c_{\theta}\right. \nonumber \\
 && +A_5 s^2_{\theta}s_{2\phi} +A_6 s_{2\theta}s_{\phi} +A_7
    s_{\theta}s_{\phi}     \nonumber \\
 &&+ B_3^e s_{\theta}^3
     c_{\phi}  + B_3^o s_{\theta}^3s_{\phi} +B_2^e s_{\theta}^2
     c_{\theta} c_{2\phi}  \nonumber \\
 && +B_2^o s_{\theta}^2
     c_{\theta} s_{2\phi} +\frac{B_1^e }{2}
    s_{\theta}(5c_{\theta}^2-1) c_{\phi}  \\ \nonumber
  && \left.
  +  \frac{B_1^o }{2} s_{\theta}(5c_{\theta}^2-1) s_{\phi}  + \frac{B_0}{2} (5 c_{\theta}^3-3c_{\theta})
     \right\}.
     \label{eq:newCSexp}
\end{eqnarray}
We have used the combinations of spherical harmonics
\begin{equation}
Y_3^0, \;\;Y_3^{1} \pm Y_3^{-1},
\;\; Y_3^{2} \pm Y_3^{-2},\;\; Y_3^{3} \pm Y_3^{-3}.
\end{equation}
in forming the basis for the new $B_i^{e,o}$ coefficients.  The superscripts $e,o$ on the new $B_i$ coefficients refer to either
even or odd under T-reversal.  The
amplitude of Eq.~(\ref{eq:CLLd3amp}) populates the $B_0$ coefficient.
The $B_i^{o,e}$ coefficients with $i>0$ are first populated at ${\cal
  O}(\alpha_s)$.

We present here representative numerical results 
to assess the potential observation of these effects.  We assume
$\sqrt{s}=14$ TeV collisions.  Our hadronic results use the NNPDF 3.1
parton distribution functions extracted to NLO precision~\cite{NNPDF:2017mvq}, and
assume an on-shell electroweak scheme with $G_{\mu}$, $M_W$, and
$M_Z$ taken as input parameters.  We impose the following cut on the invariant mass of the
final-state system: $m_{ll}>100$ GeV. We focus on the $B_0$ coefficient here.  We set the renormalization and
factorization scales to $\mu = m_{ll}$. While the $B_i$ are not generated in the SM from
perturbative QCD corrections, they can be obtained from higher-order
electroweak effects.  The leading contributions to the $B_0$
coefficient are the angular-dependent next-to-leading logarithmic (NLL)
electroweak Sudakov logarithms.  The leading logarithms depend only on
the Mandelstam invariant $\hat{s}$, and therefore do not induce any
$B_i$ coefficients.  We study the leading one-loop NLL
electroweak Sudakov logarithms in the SM using the results of
Ref.~\cite{Denner:2006jr}.

\begin{figure}[h!]
\centering
\includegraphics[width=0.9\textwidth]{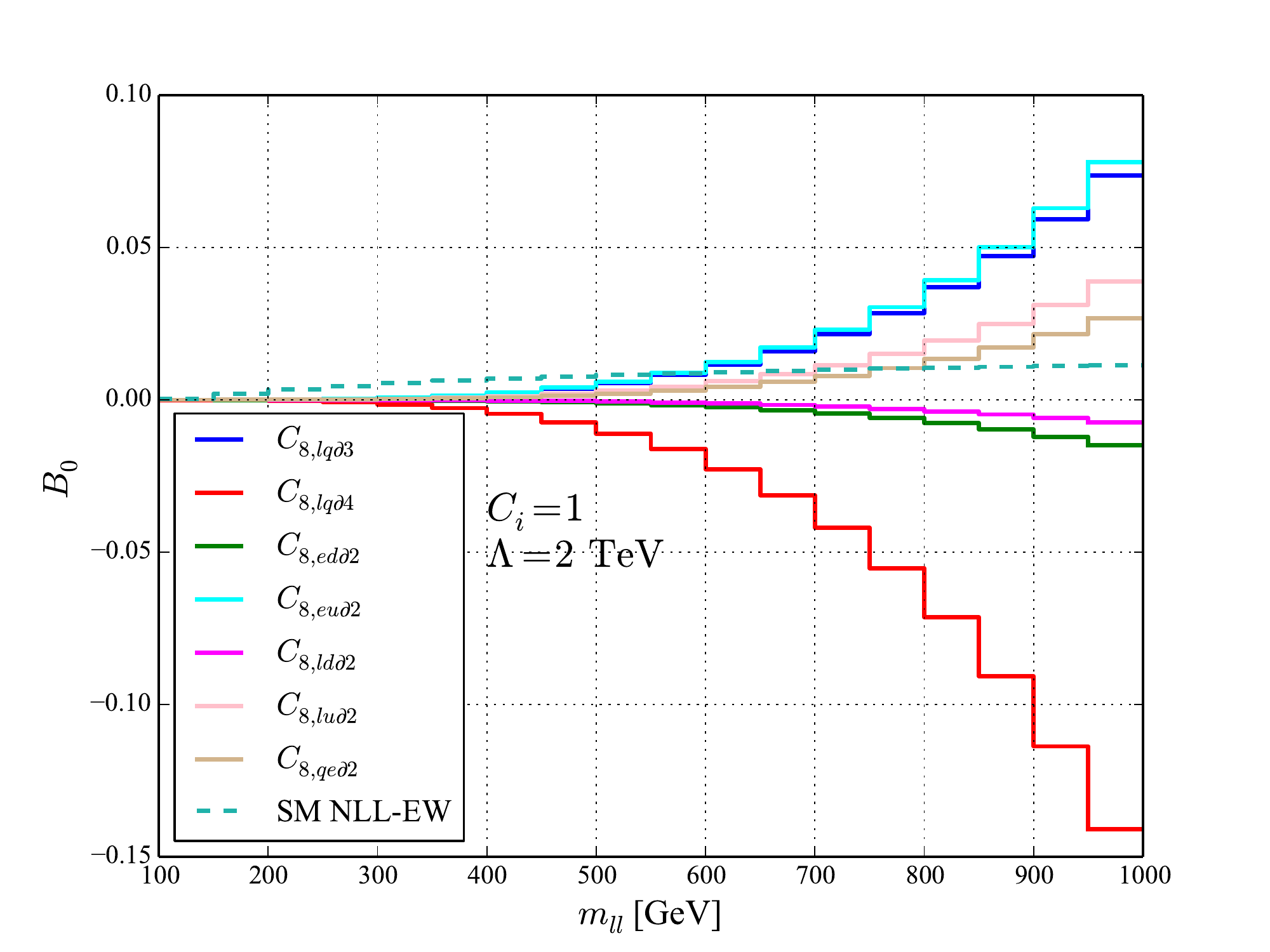}
\caption{$B_0$ coefficient as a function of the dilepton invariant mass.
\label{fig:B0}}
\end{figure}

We show in Fig.~\ref{fig:B0} $B_0$ as a function
of the invariant mass $m_{ll}$ for the seven contributing operators.  We
set $\Lambda = 2\, {\rm TeV}$  and each
Wilson coefficient separately to $C_{i} = 1$ while
setting the others to zero to obtain these seven curves.
The SM contribution is small since
it grows only logarithmically with invariant mass as ${\rm
  log}(m_{ll}/M_Z)$.   The
dimension-8 contributions grow polynomially as $m_{ll}^4$, as can be seen from the example matrix element in Eq.~(\ref{eq:CLLd3amp}) upon
setting $\hat{s}=m_{ll}^2$. The SMEFT-induced corrections are clearly visible over the SM contribution. We note that we have
calculated the corrections quadratic in the dimension-8
coefficients. The total correction from dimension-8 operators is at most 30\% of the SM.
$B_0$ receives corrections of similar size.  We therefore conclude that
the linear dimension-8 terms contribute the dominant correction to
both the cross section and the $B_0$ angular coefficient in the
invariant mass region considered, and that the truncation of the EFT
expansion to the linear dimension-8 level is justified in our study.

\begin{figure}[h!]
\centering
\includegraphics[width=0.8\textwidth]{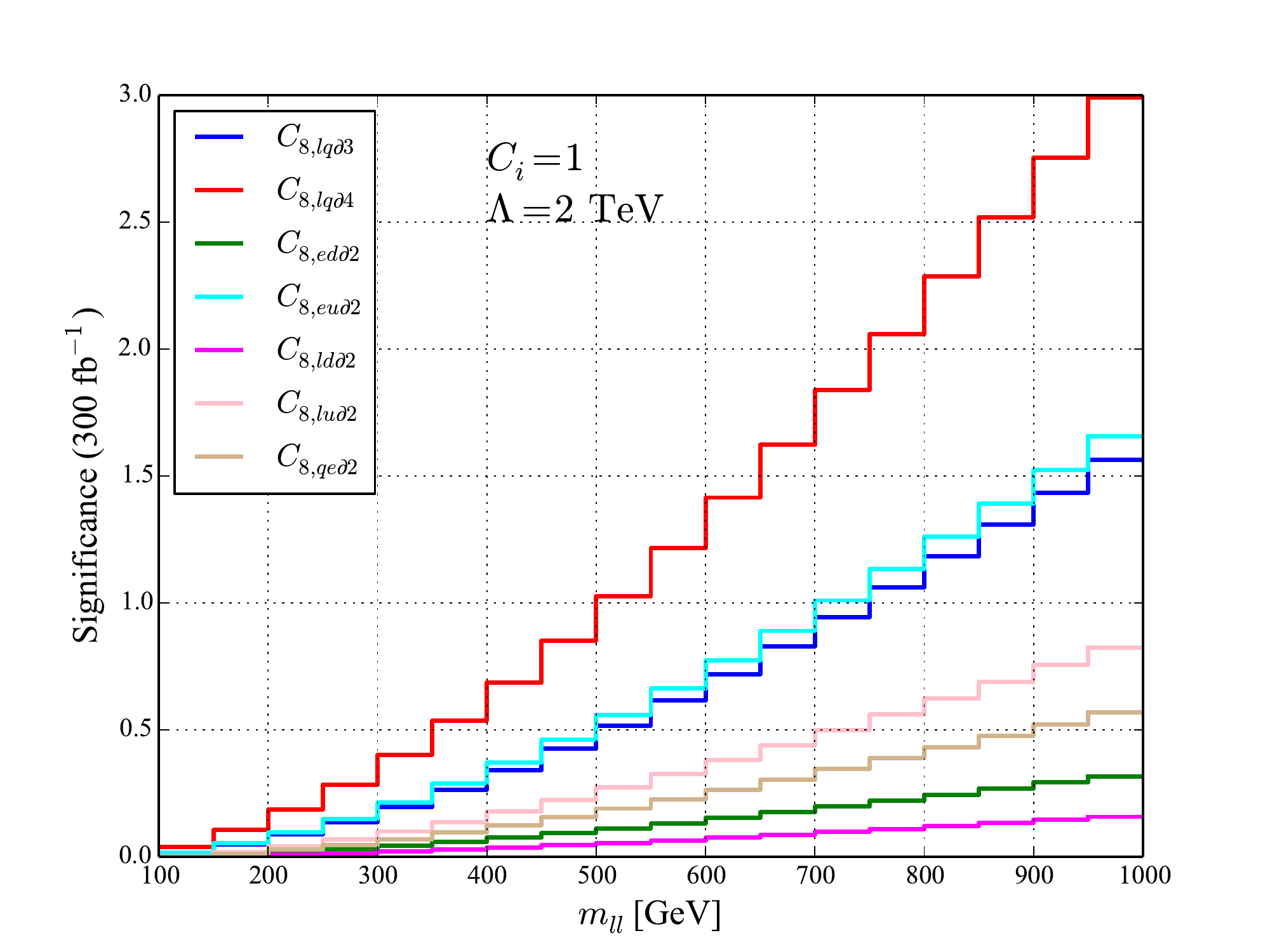}
\caption{Statistical significance of the $B_0$ angular dependence as a function of the dilepton invariant mass.
\label{fig:Sig}}
\end{figure}

We next estimate the sensitivity of the LHC to this effect by applying the optimal observable technique~\cite{Atwood:1991ka} as described in Ref.~\cite{Alioli:2020kez}. We show in Fig.~\ref{fig:Sig} the statistical significance of as a function of dilepton invariant mass for each
of the dimension-8 coefficients assuming 300 fb$^{-1}$ of integrated
luminosity.
The statistical significance for 3000 fb$^{-1}$, the target of the High Luminosity LHC (HL-LHC), is obtained by rescaling Fig.~\ref{fig:Sig} by $\sqrt{10}$.
We see that the statistical
significance per bin reaches 3 for the $C_{8,lq\partial 4}$
coefficient at high invariant mass, while for  $C_{8,lq\partial 3}$
and $C_{8,eu\partial 2}$ it reaches 1.5.  
This indicates that the
effects of $C_{8,lq\partial 4}$ should be significantly larger than statistical
fluctuations in the data at the LHC Run 3. For $\Lambda = 2$ TeV, all three coefficients 
should be visible at the HL-LHC. The statistical significance is further increased by considering correlations between different invariant mass bins.
If we combine all bins between 650 and 1000 GeV, the significance with 300 fb$^{-1}$ of integrated luminosity
reaches more than 6 for $C_{8,lq\partial 4}$, more than 3.5 for 
$C_{8,lq\partial 3}$ and $C_{8,eu\partial 2}$, and more than 1.5 for $C_{8, lu \partial 2}$.
Searches for the $B_0$ coefficient
at the future LHC are therefore promising.

\subsection{Bounds on dimension-6 Wilson coefficients including dimension-8 effects}\label{bound}

We now address the interplay between dimension-6 and dimension-8 operators in fits to LHC Drell-Yan data. As many fits to the available data consider only dimension-6 operators it is important to determine the stability of this framework under the addition of higher-dimension effects. We address this by considering the four-fermion sector of the SMEFT. As shown in Ref.~\cite{Boughezal:2021tih}, other corrections at dimension-8 such as fermion-boson vertex corrections cannot be probed with the current data precision. The following Lagrangian describes the seven contributing dimension-6 operators:
\begin{eqnarray}\label{eq:fourfermion}
\mathcal L_{\psi^4} &=& \frac{1}{\Lambda^2} \bigg\{ 
C^{(1)}_{\ell q}\, \bar \ell_L \gamma^\mu \ell_L \, \bar q_L \gamma_\mu q_L + C^{(3)}_{\ell q} \, \bar \ell_L \tau^I \gamma^\mu \ell_L \, \bar q_L  \tau^I \gamma_\mu q_L \nonumber \\
& & + C_{eu} \, \bar e_R \gamma^\mu e_R \, \bar u_R \gamma_\mu u_R  +\
 C_{ed} \, \bar e_R \gamma^\mu e_R \, \bar d_R \gamma_\mu d_R \nonumber \\
 & & + C_{\ell u}\, \bar \ell_L \gamma^\mu \ell_L \, \bar u_R \gamma_\mu u_R +  C_{\ell d}\, \bar \ell_L \gamma^\mu \ell_L \, \bar d_R \gamma_\mu d_R  +\ C_{q e}  \, \bar e_R \gamma^\mu e_R \, \bar q_L \gamma_\mu q_L   \bigg\}.
\end{eqnarray}
In this analysis we study only vector four-fermion operators. More details on scalar and tensor operators, which first contribute at ${\cal O}(1/\Lambda^4)$, can be found in~\cite{Boughezal:2021tih}. The relevant dimension-8 operators contributing to inclusive observables such as the Drell-Yan invariant mass or transverse momentum distributions are described by the following Lagrangian:
\begin{eqnarray}\label{eq:fourfermion8}
\mathcal L_{\psi^4 D^2} &=& \frac{1}{\Lambda^4} \bigg\{ 
C^{(1)}_{\ell^2\, q^2\,  D^2}\, \partial_\nu \left( \bar \ell_L \gamma^\mu \ell_L\right) \, \partial^\nu \left(\bar q_L \gamma_\mu q_L\right) + 
C^{(3)}_{\ell^2\, q^2\,  D^2}\, D_\nu \left( \bar \ell_L \gamma^\mu \tau^I \ell_L\right) \, D^\nu \left(\bar q_L \gamma_\mu \tau^I q_L\right) \nonumber
\\
& & + C^{(1)}_{e^2\,u^2\, D^2} \, \partial_\nu \left(\bar e_R \gamma^\mu e_R \right)\, \partial^\nu \left( \bar u_R \gamma_\mu u_R \right) +\
 C^{(1)}_{e^2\,d^2\, D^2} \, \partial_\nu \left( \bar e_R \gamma^\mu e_R \right) \, \partial^\nu \left( \bar d_R \gamma_\mu d_R \right) \nonumber \\
 & & + C^{(1)}_{ \ell^2\,u^2\, D^2}\, \partial_\nu \left( \bar \ell_L \gamma^\mu \ell_L \right) \, \partial^\nu \left( \bar u_R \gamma_\mu u_R\right) +  C^{(1)}_{\ell^2\, d^2\, D^2}\, \partial_\nu \left( \bar \ell_L \gamma^\mu \ell_L\right) \, \partial^\nu \left( \bar d_R \gamma_\mu d_R\right) \nonumber \\ & & +\ C^{(1)}_{ q^2\, e^2\, D^2}  \, \partial_\nu \left( \bar e_R \gamma^\mu e_R\right) \, \partial^\nu \left( \bar q_L \gamma_\mu q_L \right)  \bigg\}.  
\end{eqnarray}
These seven operators appear in exact analogy to the dimension-6 ones that appear in Eq.~(\ref{eq:fourfermion}), only with two additional derivatives. We note that the operators that affect Drell-Yan angular distributions as discussed in the previous subsection can be arranged to vanish upon integration over angles, and therefore do not contribute to the invariant mass or transverse momentum distributions~\cite{Boughezal:2021tih}. The operators listed above give rise to shifts of the cross section that scale as ${\cal O}(\hat{s}^2/\Lambda^4)$ and become important in  high invariant mass bins being probed by current LHC data. We will see this clearly in the numerical analysis that follows.

We now extract bounds on SMEFT coefficients from the  results of Ref. \cite{ATLAS:2016gic}, which measured $p p \rightarrow \ell^+ \ell^-$, with $\ell = \{e,\mu\}$ at 8 TeV with luminosity $20.3$ fb$^{-1}$. The data are binned in twelve invariant mass bins with $m_{\ell \ell}$ varying between $m_{\ell \ell} = 116$ GeV and $m_{\ell \ell} = 1.5$ TeV. The experimental uncertainties go from 0.63\% in the smallest invariant mass bin to $17.31\%$ in the highest invariant mass bin. The uncertainty in the lower invariant mass bins is an approximately equal split between statistical and systematic errors, while in the highest bins it is dominated by statistics. One important
feature of the data set of Ref.~\cite{ATLAS:2016gic} is that it was originally
intended as a SM measurement, and  a
careful accounting of experimental errors was performed and released publicly. This is an
important point that can outweigh the improvement in constraints expected from  higher-energy LHC collisions if those are not done with the same level of detail. 

We choose the UV scale $\Lambda = 4$ TeV, which is above the highest invariant mass bin studied in the experimental analysis. We calculate the SM cross section at next-to-next-to leading order (N$^2$LO) in QCD and include next-to-leading-logarithmic (NLL) electroweak corrections. The SMEFT-induced corrections are calculated at NLO in the QCD coupling constant. We have assumed no underlying hierarchy regarding the dimension-6 and dimension-8 coefficients, and rely instead upon the experimental data to determine their allowed ranges. For more details on our calculational procedure we refer to Ref.~\cite{Boughezal:2021tih}.

\begin{figure}
\begin{centering}
 \includegraphics[width=0.7\textwidth]{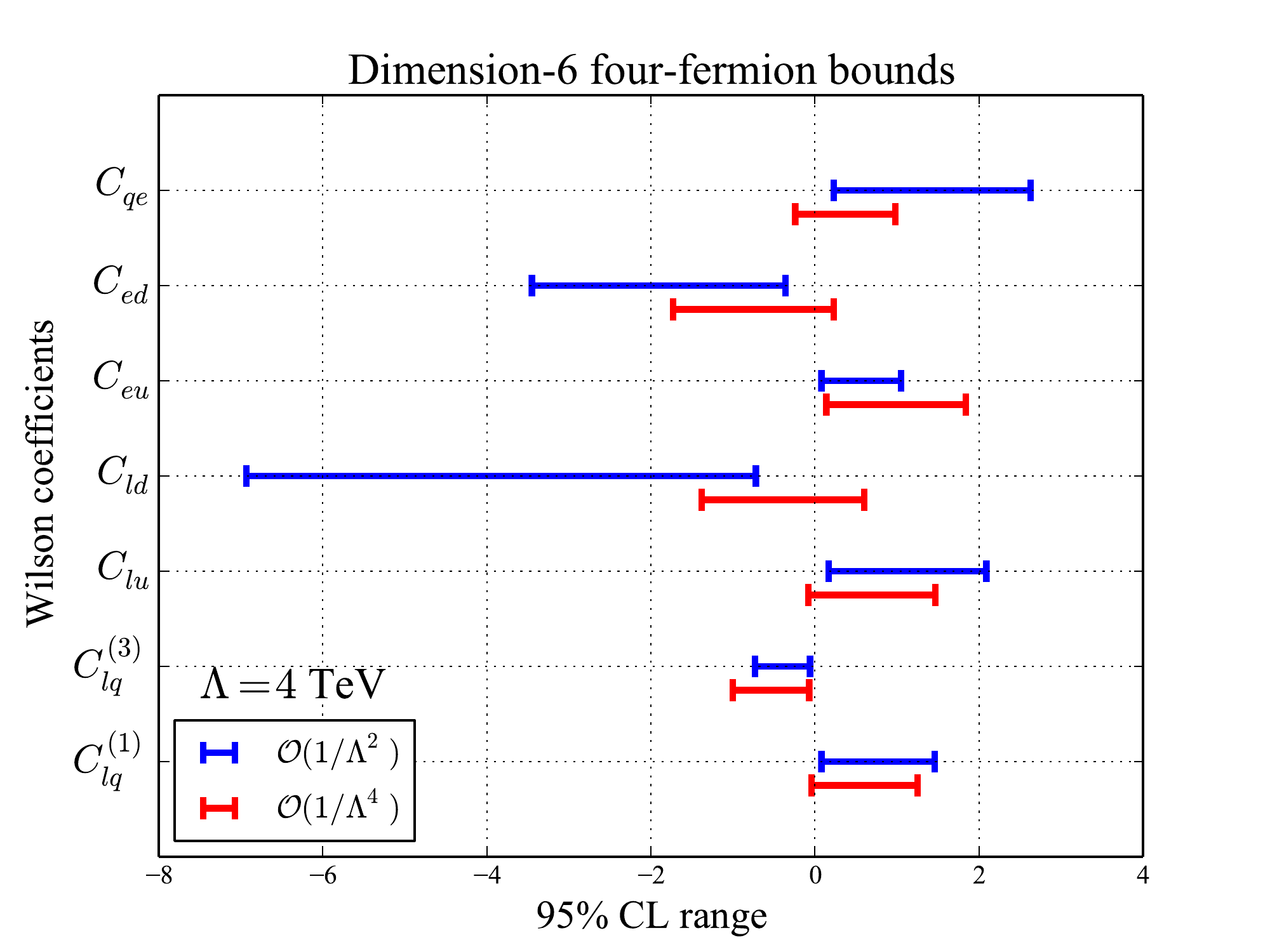}
 \caption{95\% CL intervals for the dimension-6 four-fermion operators that interfere with the SM. Both the limits obtained by considering only 
$\mathcal O(1/\Lambda^2)$ effects, as well as those including the $\mathcal O(1/\Lambda^4)$ corrections, are shown.}\label{fig:bound1}
\end{centering}
\end{figure}

 We first begin by discussing the bounds on dimension-6 four-fermion coefficients in Fig.~\ref{fig:bound1}. We compare the results obtained by keeping only $\mathcal O(1 /\Lambda^2)$ corrections with those obtained by keeping the square of dimension-6 operators that contributes $\mathcal O(1 /\Lambda^4)$ effects. In general the fits to the data are good, with a $\chi^2/$dof below one for most operators, implying that the data prefer a non-zero contribution from SMEFT operators, which interfere destructively with the SM.
The impact of $\mathcal O(1 /\Lambda^4)$ corrections are significant for most operators, with the upper and lower limits of the 95\% CL ranges shifting by factors of 2 or 3 in most cases.

\begin{figure}
\begin{centering}
 \includegraphics[width=0.7\textwidth]{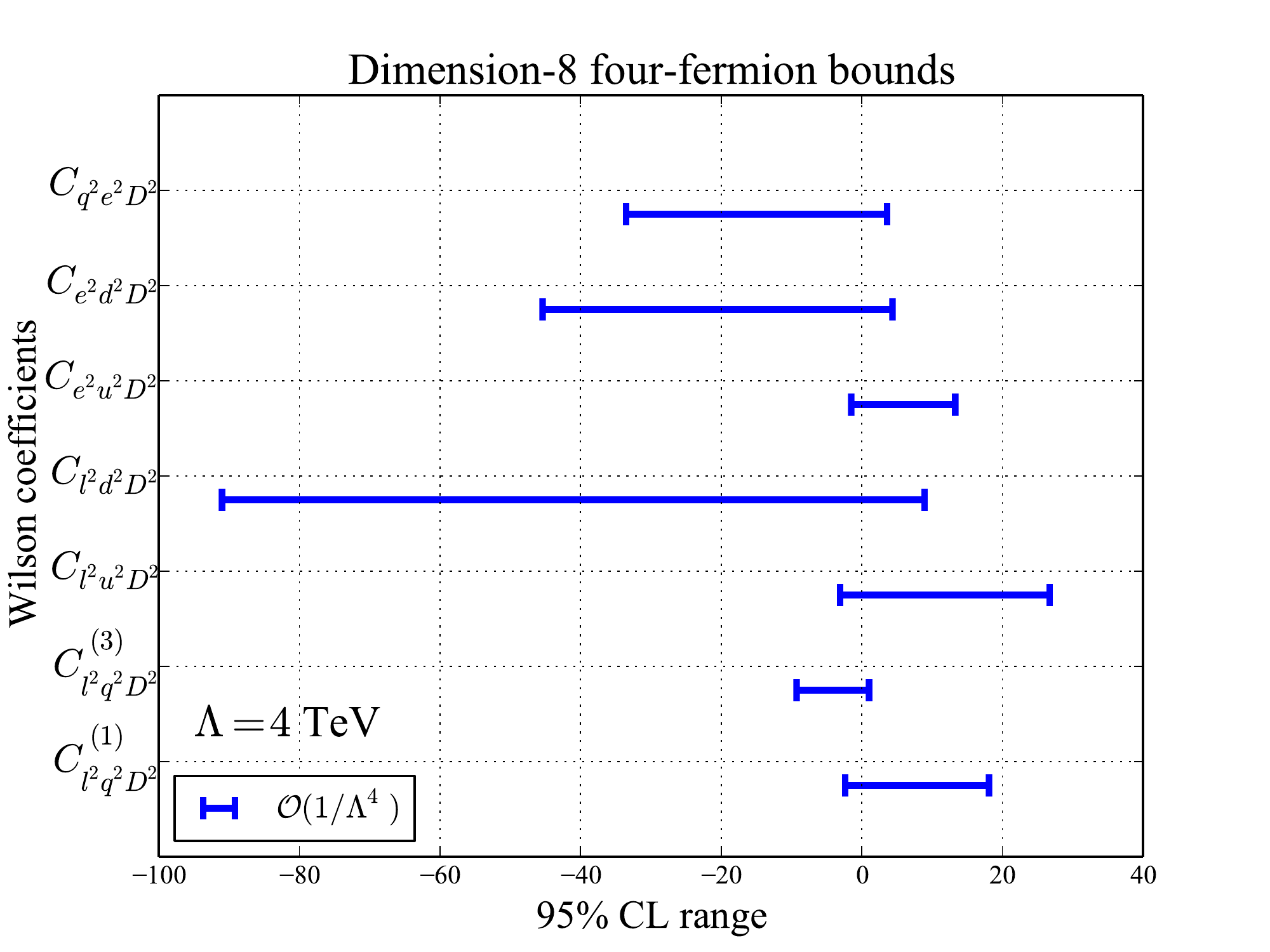}
 \caption{95\% CL intervals for the dimension-8 momentum-dependent four-fermion operators. }\label{fig:bound2}
\end{centering}
\end{figure}

We next proceed to discuss the dimension-8 operator bounds in Fig.~\ref{fig:bound2}. Defining the effective scale as $\Lambda/\sqrt[\leftroot{-2}\uproot{2}4]{C_i}$ for these dimension-8 terms, UV scales ranging from 1.3 to 4.3~TeV are reached for operators of the $\mathcal L_{\psi^4 D^2}$ class. Given that these are pure dimension-8 effects that appear first at $\mathcal O(1/\Lambda^4)$ this is striking. Since these scales approach those of the dimension-6 effects, these operators cannot be safely neglected in fits to the data.

In this section we have focused on the phenomenlogical consequences of dimension-6 and dimension-8 operators in the Drell-Yan process. By itself, the Drell-Yan process suffers from degeneracies between Wilson coefficients due to the limited number of observables that can be measured. Lifting these degeneracies requires the consideration of other experiments, such as a future electron-ion collider~\cite{Boughezal:2020uwq}. A combination of low-energy observables with Drell-Yan data can help remove ambiguities between dimension-6 and dimension-8 effects that occur at high invariant masses in Drell-Yan~\cite{Boughezal:2021kla}.

\subsection{Dimension-8 effects in low-energy precision experiments}

Low-energy precision experiments can provide very strong bounds on SMEFT operators, competitive and complementary to high-energy colliders. This is true in particular for observables sensitive to the violation of approximate SM symmetries. Constraints on  electric dipole moments of leptons, nucleons,
atoms and molecules, on the neutrinoless double beta half-life of $^{76}$Ge, $^{130}$Te and $^{136}$Xe, and on $\mu \rightarrow e$ transitions, probe flavor-diagonal CP-violation, lepton-number violation and charged-lepton-flavor violation at levels that are (naively) out of the LHC's reach.  
A similar argument applies to flavor observables, including  meson-antimeson oscillations, rare $B$ meson decays or direct CP-violation in kaon decays.
In other cases, low-energy experiments and colliders complement each other. This  for example applies to  non-standard charged-current interactions, for which Drell-Yan experiments and $\beta$ decays probe similar parameter space \cite{Cirigliano:2012ab,Alioli:2018ljm,Falkowski:2020pma}, and Drell-Yan data can be used to falsify explanations of low-energy anomalies \cite{Alok:2021ydy}. 

Dimension-6 operators that induce semileptonic charged-current processes are probed by tests of unitarity of the CKM matrix, by pion, kaon,  neutron and nuclear $\beta$ decay rates and correlation coefficients, and by precise measurements of the $\beta$ spectra. The bounds on their coefficients are typically  $\mathcal O(10^{-3})$, in units of the Fermi constant $G_F$, which naively corresponds to scales of 5 to 10 TeV. For comprehensive analyses of existing low-energy constraints, we refer to Refs. \cite{Cirigliano:2013xha,Falkowski:2020pma,Cirigliano:2021yto}.
To test the sensitivity of $\beta$ decay experiments to dimension-8 operators, we can consider the two derivative operators $C^{(3)}_{\ell^2\, q^2\,  D^2}$ and $C^{(4)}_{\ell^2\, q^2\,  D^2}$
\begin{eqnarray}\label{beta8}
    \mathcal L &=& \frac{1}{\Lambda^4} 
    C^{(3)}_{\ell^2\, q^2\,  D^2}\, D_\nu \left( \bar \ell_L \gamma^\mu \tau^I \ell_L\right) \, D^\nu \left(\bar q_L \gamma_\mu \tau^I q_L\right) \nonumber \\ && +
\frac{1}{\Lambda^4}     C^{(4)}_{\ell^2\, q^2\,  D^2}\, \left( \bar \ell_L \gamma^{\{\mu} \overleftrightarrow{D}^{\nu\}} \tau^I \ell_L\right) \, D^\nu \left(\bar q_L \gamma_{\{\mu} \overleftrightarrow{D}_{\nu\}} \tau^I q_L\right), 
 \end{eqnarray}
where $\{ .\, , . \}$ denotes the symmetric, traceless combination. 
We slightly redefined $C^{(4)}_{\ell^2\, q^2\,  D^2}$ with respect to Section \ref{angular} to simplify its nucleon matrix element.
Focusing on neutron decay, the matrix element of $C^{(3)}_{\ell^2\, q^2\,  D^2}$ can be 
simply expressed in terms of the neutron axial and vector current, while the matrix element of 
$C^{(4)}_{\ell^2\, q^2\,  D^2}$
is related to moments of the parton distributions 
\begin{eqnarray}\label{nme}
\langle p | \bar q \tau^+ \gamma^{\mu} P_L q  | n \rangle &=& \bar u^p (v^\mu - 2 g_A S^\mu) u^n\\ 
    \langle p | \bar q \tau^+ \gamma^{\{\mu} \overleftrightarrow D^{\nu\}} P_L q  | n \rangle &=&  \frac{1}{2} m_N \bar u^p \Bigg\{ \langle x \rangle_{u-d}  \left( v^{\mu} v^\nu - \frac{1}{4} g^{\mu \nu} \right)  
     + 2 \langle x \rangle_{\Delta u- \Delta d}    v^{\{\mu} S^{\nu\}}    \Bigg\} u^n,\nonumber \\
\end{eqnarray}
where $m_N$ ,
$v^\mu$ and $S^\mu$ are the nucleon mass, velocity and spin.
$\langle x \rangle_{u-d}$ and $\langle x \rangle_{\Delta u- \Delta d}$ have been computed in Lattice QCD,
e.g.  $\langle x \rangle_{u-d} = 0.173(14)$
and $\langle x \rangle_{\Delta u- \Delta d} = 0.225(22)$ \cite{Mondal:2020cmt}, or can be extracted from PDF fits. 

Using the matrix elements in Eq. \eqref{nme}, we can derive the corrections to the neutron decay rate, differential in the electron energy and in the angle between the momenta of electron and neutrino $\theta$: 
\begin{eqnarray}
 & &\frac{d \Gamma}{d E_e d \cos\theta} = \frac{(G_F V_{ud})^2}{4\pi^3} Ee |\vec p_e|
(E_0 - E_e)^2 (1+ 3 g_A^2)
\left\{ 1 + c_\theta \frac{1 - g_A^2}{1+3 g_A^2} \right. \nonumber \\ & &  \left.- \frac{2 v^2}{\Lambda^4}C^{(3)}_{\ell^2\, q^2\,  D^2} \left( 2 E_e (E_0 -E_e) + m_e^2\right) \left(1 - \frac{1+ 7 g_A^2}{1 + 3 g_A^2} c_\theta
-  \frac{1 - 5 g_A^2}{1 + 3 g_A^2} c_\theta^2 + \frac{1-g_A^2}{1+3 g_A^2} c_\theta^3
\right)  \right. \nonumber \\
 & & \left. - \frac{2 v^2 m_N}{\Lambda^4}C^{(4)}_{\ell^2\, q^2\,  D^2} m_N \frac{2 E_e - E_0}{ 3 g_A^2+1} \Bigg( ( 8 g_A \langle x \rangle_{\Delta u- \Delta d}  - \langle x \rangle_{ u-  d}  )
 + c_\theta \langle x \rangle_{ u-  d}   \Bigg) + \ldots
 \right\},
\end{eqnarray}
where $c_\theta \equiv \cos\theta$,
$E_0$ is the electron end-point energy,
and $\ldots$ denotes terms proportional to the electron mass, which we omitted for simplicity.
The dimension-8 operators induce new energy and angular dependencies, e.g. they induce terms proportional to $c^2_\theta $, which in the SM only arise at $\mathcal O(E_e/m_N)$. 
Still, the corrections to neutron correlation coefficients scale at best as 
\begin{equation}
 \frac{ c_{\rm dim 8}}{c_{\rm SM}}\sim     \frac{m_N^2}{\Lambda^2} \frac{v^2}{\Lambda^2},
\end{equation}
for those coefficients that are suppressed in the SM. Even if $\Lambda \sim v$, these corrections are too small to be observed in the next generation of $\beta$ decay experiments.
Differently from the LHC analysis discussed in Section \ref{bound}, low-energy experiments are more sensitive to dimension-8 operators with insertion of Higgs fields. In these cases, the corrections to neutron decay scale as $(v/\Lambda)^4$ and current low-energy charged-current data are sensitive to dimension-8 operators with $\Lambda \sim 1.5$ TeV. A joint analysis of collider and low-energy probes of charged-current interactions to dimension-8 would therefore be an important development.

Another interesting class of low-energy observables which can show sensitivity to dimension-8 operators is electric dipole moments (EDMs). Also in this case, we can differentiate between dimension-8 operators that contain insertions of Higgs fields vs operators with additional derivatives or gauge fields.
Even for dimension-6 operators, the evaluation of hadronic EDMs is complicated by non-perturbative QCD and an open field of research \cite{Shindler:2021bcx}. Using only naive dimensional analysis (NDA), we can estimate the contribution to the neutron EDM to scale as
\begin{equation}
    d_n \sim \frac{e}{(4\pi)^2} \left\{ \frac{\Lambda_\chi}{v^2 }, 
\frac{\Lambda^3_\chi}{v^4}     \right\}\frac{v^4}{\Lambda^4} \sim  \left\{ 10^{-8}, 10^{-13} \right\} \left(\frac{v}{\Lambda}\right)^4 \, e \, {\rm fm},
\end{equation}
with $\Lambda_\chi \sim m_N$.
The first estimate applies, for example, to four-quark two-Higgs operators, while the second to operators with four quarks and one gluon field. Comparing the NDA estimate with the current bound on the neutron EDM, $d_n < 1.8 \cdot 10^{-13}$ $e$ fm, we see that EDMs can probe at least some classes of dimension-8 operators.

Finally, at dimension-8, a new class of flavor diagonal CP-odd operators that break time-reversal, but not parity, arises. These generate 
new $T$-odd $P$-even effects at low energy, including
toroidal quadrupole moments of particles with angular momentum greater than one (such as the deuteron or positronium) \cite{PhysRevA.49.5105,Mereghetti:2013bta}
and $T$-violating asymmetries in proton-deuteron scattering \cite{Aksentyev:2017dnk}. 
Working towards a global SMEFT fit, it will be important to assess the required sensitivity
for these experiments to provide competitive bounds on BSM physics.


\section{Conclusions} \label{sec:conc}

The search for physics beyond the Standard Model is increasingly pushing the allowed scale for new particles beyond the energy reach of current colliders. The SMEFT is a systematic framework for exploring the impact of such high-energy states in lower-energy measurements. It relies on an expansion in inverse powers of a heavy scale $\Lambda$. Like most other expansions in physics, reliable predictions require going beyond the first non-trivial orderized by dimension-6 operators. 

In this contribution we have reviewed the theoretical framework that has allowed predictions at the dimension-8 level and beyond in the SMEFT. These advances have required a wide variety of theoretical techniques, that in some cases have led to predictions to all orders in the $1/\Lambda$ expansion. The phenomenological consequences of these higher-order terms are significant in several benchmark processes. Further investigation of these effects will be needed as we enter the high-luminosity stage of the LHC and prepare for future high precision experiments.

\section{Acknowledgements}  \label{sec:acc}

R.~B. is supported by the DOE contract DE-AC02-06CH11357.
S.D.B is supported by SRA (Spain) under Grant No.\
PID2019-106087GB-C21/10.13039/501100011033 as well as 
by the Junta de Andaluc\'ia (Spain) under Grants No.\ FQM-101,
A-FQM-467-UGR18, and P18-FR-4314 (FEDER).
W.C.\ is supported by  the Global Science Graduate Course (GSGC) program of the University of Tokyo, the World Premier International Research Center Initiative (WPI) and acknowledges support from JSPS KAKENHI grant number JP19H05810. 
M.C. is supported by the Spanish MINECO under the Ram\'on y Cajal programme as well as by Junta de Andaluc\'ia (Spain) under grants FQM-101 and A-FQM-211-UGR18. 
L.G. acknowledges support from the National Science Foundation, Grant PHY-1630782, and to the Heising-Simons Foundation, Grant 2017-228.
G.G is supported by LIP (FCT, COMPETE2020-Portugal2020, FEDER, POCI-01-0145-FEDER-007334) as well as by INCD under the project CPCA-A1-401197-2021 and by FCT under the project CERN/FIS-PAR/0024/2019 and under the grant SFRH/BD/144244/2019.
X.~Lu is supported by the DOE grant DE-SC0011640.
C.S.M.\ is supported by the Deutsche Forschungsgemeinschaft under Germany's Excellence Strategy  EXC 2121 ``Quantum Universe'' - 390833306.
A.M.\ is supported by the National Science Foundation under grant number PHY-2112540.
T.M.\ is supported by the World Premier International Research Center Initiative (WPI) MEXT, Japan, and by JSPS KAKENHI grants JP18K13533, JP19H05810, JP20H01896, and JP20H00153. 
E.~M. is supported  by the US Department of Energy through  
the Office of Nuclear Physics  and  the  
LDRD program at Los Alamos National Laboratory. Los Alamos National Laboratory is operated by Triad National Security, LLC, for the National Nuclear Security Administration of U.S.\ Department of Energy (Contract No. 89233218CNA000001).
J.R.N.\ is supported by the Deutsche Forschungsgemeinschaft (DFG, German Research Foundation) - Projektnummer 417533893/GRK2575 “Rethinking Quantum Field Theory”.
S.P.\ acknowledges funding provided
by Tomislav and Vesna Kundic as well as the support from DOE grant DE-SC0009988. 
F.~P.\ is supported by the DOE grants DE-FG02-91ER40684 and DE-AC02-06CH11357.
J.S. is supported by the National Natural Science Foundation of China under Grants No.12025507, No.12150015, No.12047503; and is supported by the Strategic Priority Research Program and Key Research Program of Frontier Science of the Chinese Academy of Sciences under Grants No. XDB21010200, No. XDB23010000, and No. ZDBS-LY-7003 and CAS project for Young Scientists in Basic Research YSBR-006.

\bibliographystyle{apsrev4-1_title}
\bibliography{main}
\end{document}